\documentclass{article}
\usepackage[left=1in, right=1in, bottom = 1in, top = 1in]{geometry}
\usepackage{graphicx} % Required for inserting images
\usepackage{bbm}
\usepackage{amsmath}
\usepackage[english]{babel}
\usepackage[utf8]{inputenc}
\usepackage[authoryear]{natbib}
\usepackage{setspace}

\usepackage{amsmath}
\usepackage{amsfonts}
\usepackage{amsthm}
\usepackage{amssymb}
\usepackage{mathrsfs}
\usepackage{mathtools}
\usepackage{amstext}
\usepackage{comment}
\usepackage{epsfig}
\usepackage{latexsym}
\usepackage{epstopdf}
\usepackage{bm}
\usepackage{algorithm}
\usepackage{algpseudocode}
\usepackage{tikz}
\usetikzlibrary{positioning,chains,fit,shapes,calc}
\definecolor{myblue}{RGB}{80,80,160}
\definecolor{mygreen}{RGB}{80,160,80}

\usepackage{url}
\usepackage[colorlinks = true, linkcolor = blue]{hyperref}
\hypersetup{
        urlcolor = blue,
	colorlinks = true,
	linkcolor = blue,
	citecolor = blue,
}
\usepackage{paralist}
\usepackage{enumitem}
\setlist[enumerate]{topsep=0pt,noitemsep}

\usepackage{multirow}
\usepackage{multicol}
\usepackage{babel}
\usepackage{array}
\usepackage{rotating}
\usepackage{longtable}
\usepackage{float}
\usepackage{booktabs}
\usepackage{lscape}
\usepackage{caption}
\usepackage{subcaption}
\usepackage{afterpage}
\usepackage{authblk}

\newtheorem{theorem}{Theorem}[section]
\newtheorem{remark}{Remark}[section]
\newtheorem{assumption}{Assumption}
\newtheorem{lemma}{Lemma}[section]

\newtheorem{example}{Example}
\newtheorem{definition}{Definition}[section]
\newtheorem{proposition}{Proposition}[section]

\newenvironment{continuance}[1]
{%
  \newcommand{\continuanceref}{\ref{#1}}% Define reference to original example
  \continuancex % Begin the non-numbered theorem-like environment
}
{\endcontinuancex}

\newtheorem*{continuancex}{Example \continuanceref \ (continued)}

\allowdisplaybreaks

\newcommand{\B}{B}

\def\T{{\textsc{t}}}

\newcommand{\cov}{\textup{cov}}
\newcommand{\var}{\textup{var}}
\newcommand{\avar}{\textup{avar}}
\newcommand{\acov}{\textup{acov}}

\newcommand{\bOne}{\boldsymbol{1}_{m}}
\newcommand{\bZero}{\boldsymbol{0}_{m}}
\newcommand{\Yt}[1]{Y_{#1}(\boldsymbol{1}_{m})}
\newcommand{\Yc}[1]{Y_{#1}(\boldsymbol{0}_{m})}

\newcommand{\bY}{\boldsymbol{Y}}
\newcommand{\bz}{\boldsymbol{z}}
\newcommand{\bZ}{\boldsymbol{Z}}

\newcommand{\bX}{\boldsymbol{X}}

\newcommand{\cN}{\mathcal{N}}
\newcommand{\cF}{\mathscr{F}}
\newcommand{\cT}{\mathcal{T}}
\newcommand{\cS}{\mathcal{S}}

\newcommand{\allt}{\left(\prod_{k\in \mathcal{G}_{+i}} Z_k\right)}
\newcommand{\alltj}{\left(\prod_{k\in \mathcal{G}_{+j}} Z_k\right)}
\newcommand{\allc}{\left\{\prod_{k\in \mathcal{G}_{+i}} (1-Z_k)\right\}}
\newcommand{\allcj}{\left\{\prod_{k\in \mathcal{G}_{+j}} (1-Z_k)\right\}}

\newcommand{\indicator}{\mathbbm{1}}
\newcommand{\plim}{\textup{plim}}

\newcommand{\bG}{\boldsymbol{G}}
\newcommand{\G}{G}
\newcommand{\cG}{\mathcal{G}}
\newcommand{\cGijcap}{\cG_{+i} \cap \cG_{+j}}
\newcommand{\cGijcup}{\cG_{+i} \cup \cG_{+j}}

\newcommand{\Smax}{{\max_{1\leq i\leq n}\G_{+i}}}
\newcommand{\Dmax}{{\max_{1\leq k\leq m}\G_{k+}}}
\newcommand{\Smaxsupp}{{\bar{D}_\textup{O}}}
\newcommand{\Dmaxsupp}{{\bar{D}_\textup{I}}}

\newcommand{\sumi}{\sum_{i=1}^n}
\newcommand{\sumj}{\sum_{j=1}^n}
\newcommand{\sumk}{\sum_{k=1}^m}
\newcommand{\sumij}{\sum_{i=1}^{n} \sum_{j=1}^{n}}
\newcommand{\sumu}{\sum_{u=1}^n}
\newcommand{\sumv}{\sum_{v=1}^n}
\newcommand{\maxi}{\max_{1\leq i\leq n}}

\makeatletter
\newcommand{\ostar}{\mathbin{\mathpalette\make@circled\star}}
\newcommand{\make@circled}[2]{%
  \ooalign{$\m@th#1\smallbigcirc{#1}$\cr\hidewidth$\m@th#1#2$\hidewidth\cr}%
}
\newcommand{\smallbigcirc}[1]{%
  \vcenter{\hbox{\scalebox{0.77778}{$\m@th#1\bigcirc$}}}%
}
\makeatother

\title{Design-based causal inference in bipartite experiments}

\author[1]{Sizhu Lu\thanks{The first two authors contributed equally to this work.}}
\author[2]{Lei Shi$^{*}$}
\author[3]{Yue Fang}
\author[2]{Wenxin Zhang}
\author[1]{Peng Ding}

\affil[1]{Department of Statistics, University of California, Berkeley}
\affil[2]{Division of Biostatistics, University of California, Berkeley}
\affil[3]{School of Management and Economics, The Chinese University of Hong Kong, Shenzhen}

\begin{document}

\maketitle
\begin{abstract}
    Bipartite experiments arise in various fields, in which the treatments are randomized over one set of units, while the outcomes are measured over another separate set of units. However, existing methods often rely on strong model assumptions about the data-generating process. Under the potential outcomes formulation, we explore design-based causal inference in bipartite experiments under weak assumptions by leveraging the sparsity structure of the bipartite graph that connects the treatment units and outcome units. We make several contributions. First, we formulate the causal inference problem under the design-based framework that can account for the bipartite interference. Second, we propose a consistent point estimator for the total treatment effect, a policy-relevant parameter that measures the difference in the outcome means if all treatment units receive the treatment or control. Third, we establish a central limit theorem for the estimator and propose a conservative variance estimator for statistical inference. Fourth, we discuss a covariate adjustment strategy to enhance estimation efficiency.
\end{abstract}

\section{Introduction to bipartite experiments and bipartite interference}\label{sec::intro}

% background of bipartite experiment + examples

%% Bipartite graphs are very common in many fields, such as in online marketplace, environmental science, education, etc. Problem with cluster experiment..., benefits of bipartite experiments...

Bipartite experiments arise in various fields, including digital experimentation \citep{harshaw2023design, shi2024scalable}, environmental science \citep{zigler2021bipartite}, and public health \citep{zigler2025bipartite}. Using the terminology of \cite{zigler2025bipartite} for bipartite experiments, the treatments are randomized over one set of units, called the intervention units, while the outcomes are measured over another separate set of units, called the outcome units.
Thus, bipartite experiments are different from the classic experiments, which randomly assign units to different treatment groups and then measure the outcomes of the same set of units after treatment initiation. As illustrated in Figure~\ref{fig::illustration_bipartite_exp}, the intervention units and outcome units are connected through a known fixed bipartite graph, and the dependence of the outcome units on the intervention units is represented by the bipartite graph, leading to bipartite interference \citep{zigler2021bipartite}. In this setting, each outcome unit can possibly depend on multiple intervention units, and each intervention unit can possibly affect multiple outcome units. Because different outcome units can share connections to the same intervention units, interference naturally arises among outcome units.

% \begin{figure}[h]
%     \centering
%     \includegraphics[width=0.5\linewidth]{figures/bipartite_graph_enhanced.png}
%     \caption{Illustration of a bipartite experiment}
%     \label{fig::illustration_bipartite_exp}
% \end{figure}
\begin{figure}[!ht]
\centering
\begin{tikzpicture}[thick,
  every node/.style={draw,circle},
  fsnode/.style={fill=myblue},
  ssnode/.style={fill=mygreen},
  every fit/.style={ellipse,draw,inner sep=-2pt,text width=2cm}
]
% the vertices of U
\begin{scope}[yshift=-0.5cm,start chain=going below,node distance=7mm]
\foreach \i in {1,2,3,4}
  \node[fsnode,on chain] (u\i) [label=left: \i] {};
\end{scope}

% the vertices of V
\begin{scope}[xshift=4cm,start chain=going below,node distance=7mm]
\foreach \i in {1,2,3,4,5}
  \node[ssnode,on chain] (g\i) [label=right: \i] {};
\end{scope}

% the set U
\node [myblue,fit=(u1) (u4),label=left:intervention units] {};
% the set V
\node [mygreen,fit=(g1) (g5),label=right:outcome units] {};

% the edges
\draw (u1) -- (g1);
\draw (u2) -- (g2);
\draw (u2) -- (g3);
\draw (u3) -- (g3);
\draw (u4) -- (g1);
% \draw (u3) -- (g2);
\draw (u3) -- (g3);
\draw (u4) -- (g5);
\draw (u4) -- (g4);
\end{tikzpicture}
\caption{A bipartite experiment with $n=4$ intervention units and $m=5$ outcome units}
\label{fig::illustration_bipartite_exp}
\end{figure}

% literature review
Recent years have witnessed a growing interest in causal inference with bipartite experiments. A well-known special case is cluster randomization, which has been both theoretically studied and widely applied \citep{donner1998some, donner2000design, su2021model}. In the cluster experiment setup, each outcome unit is connected to exactly one intervention unit, while each intervention unit can be connected to multiple outcome units. More recent studies have extended this framework to general bipartite graphs, where each outcome unit may be connected to multiple intervention units, and vice versa. Several studies have developed methods for causal inference in bipartite experiments. \cite{zigler2021bipartite} introduced a formal framework for defining causal estimands in bipartite experiments and proposed an inverse probability-weighted estimator to address treatment assignment based on observed covariates. Additionally, various estimators have been developed under different model assumptions. These include approaches based on linear exposure mapping \citep{aronow2017estimating, forastiere2021identification} and linear outcome model \citep{harshaw2023design, shi2024scalable}, generalized propensity score model \citep{doudchenko2020causal}, and heterogeneous additive treatment effect model \citep{lu2024estimation}. 

% Gap in the literature
In this work, we conduct a design-based analysis of bipartite experiments, focusing on a casual parameter defined in terms of fixed potential outcomes. We derive the properties of estimators for causal effects under the randomness of treatment assignment \citep{imbens2015causal,ding2024first}. While several existing approaches \citep[e.g.][]{harshaw2023design, doudchenko2020causal, lu2024estimation} have considered such a perspective for bipartite experiments, they rely on strong modeling assumptions, making estimation and inference heavily dependent on correct model specifications. In contrast, we propose a more flexible design-based framework that imposes fewer assumptions for bipartite experiments. 
Adopting such a framework presents several challenges. One key difficulty is to establish central limit theorems and construct valid variance estimators while accounting for dependencies across outcome units. Additionally, when covariate information is available, an important question is how to perform covariate adjustment in a model-agnostic manner without relying on parametric assumptions on the relationships between outcomes and covariates.  

We make the following contributions. 
First, we formulate the causal inference problem in bipartite experiments under the design-based framework. We generalize the classic stable unit treatment value assumption \citep{rubin1980randomization} to account for bipartite interference. This generalization is tailored to the bipartite graph structure, enabling the expression of the total treatment effect as a function of observed data. We leverage the sparsity structure of the bipartite graph to avoid the strong assumptions on modeling the potential outcomes or exposure mapping, which simplifies interference by summarizing treatment assignment through a prespecified low-dimensional vector of sufficient statistics rather than considering the full assignment. 
Second, we propose a H\'{a}jek estimator for the total treatment effect and establish its consistency and asymptotic normality. In particular, we characterize the key sparsity assumptions on the bipartite graph structure to ensure the limiting theorems. To ensure a valid inference, we also propose a conservative variance estimator that appropriately accounts for the complexity of the bipartite graph. 
Third, we discuss covariate adjustment in bipartite experiments. While covariate adjustment is a well-established technique in analyzing various experiments \citep{lin2013agnostic, liu2020regression, fogarty2018regression, su2021model, wang2024model}, interference in bipartite experiments complicates the problem. In particular, regression-based covariate-adjusted estimators may not guarantee improved estimation efficiency, and the associated robust standard errors may fail to provide valid inference without strong assumptions. We propose a model-agnostic covariate-adjusted estimator that is asymptotically at least as efficient as the H\'{a}jek-type estimator while maintaining valid inference. 

% The rest of the paper is organized as follows. Section~\ref{sec::setup} introduces the design-based setup of the bipartite experiments. Section~\ref{sec::identification-estimation} discusses the estimation of the total treatment effect under bipartite interference. Section~\ref{sec::covariate} presents a covariate adjustment strategy for constructing point and variance estimators and proves their asymptotic properties. Section~\ref{sec::simulation} conducts numerical experiments to validate the proposed methods and theoretical results. Section~\ref{sec::discussion} concludes and outlines future research directions. All proofs and technical details are in the Supplementary Material.

We will use the following notation. Let $\indicator\{\cdot\}$ denote the indicator function. Let $\bOne\in\mathbb{R}^{m}$ and $\bZero\in\mathbb{R}^{m}$ denote the vector of all 1's and 0's, respectively, and let $\boldsymbol{1}_{m\times n}\in\mathbb{R}^{m\times n}$ denote the matrix of all 1's.
Let $\asymp$ denote asymptotically the same order as $m \rightarrow \infty$. For any positive integer $K$, denote $[K]=\{1,\ldots,K\}$ as the set of all positive integers smaller than or equal to $K$. 
Write $b_m=O(a_m)$ if $b_m/a_m$ is bounded, and $b_m = o(a_m)$ if $b_m/a_m$ converges to $0$ as $m\rightarrow\infty$. Write $b_m=O_{P}(a_m)$ if $b_m/a_m$ is bounded in probability, and $b_m=o_{P}(a_m)$ if $b_m/a_m$ converges to $0$ in probability as $m\to\infty$.

\section{Setup: examples, potential outcomes, interference, and randomization}\label{sec::setup}

\subsection{Motivating examples of bipartite experiments}\label{sec::motivating-examples}
We first present three motivating examples to highlight the applicability of bipartite experiments.

\begin{example}[Power plant]\label{eg::power-plant}  
    \cite{zigler2021bipartite} studied the causal effect of installing selective noncatalytic reduction systems in upwind power plants on neighborhood hospitalization rates. The treatment is whether the installation is implemented and the intervention units are the upwind power plants. The outcome is the hospitalization rate, and the outcome units are the zip code-level areas. In their setting, a neighborhood may be affected by the treatment of multiple power plants, while a power plant may affect multiple neighborhoods. We will revisit this example in Section~\ref{sec::application}.
\end{example}

\begin{example}[Facebook Group]
    \cite{shi2024scalable} considered a bipartite experiment in Meta for testing new features on Facebook Group. The new features are randomized over Facebook user groups and outcomes are measured by user-level engagement. The outcome of each user is affected by interventions on a set of groups they belong to, while the treatment in each group affects all users within that group.
\end{example}

\begin{example}[Amazon market]\label{eg::amazon}
    \cite{harshaw2023design} studied a bipartite experiment on the Amazon marketplace to evaluate the impact of new pricing mechanisms randomized across items on the level of customer satisfaction. The treatment is the implementation of new pricing mechanisms and the intervention units are the items in the marketplace. The outcome is the satisfaction level and the outcome units are Amazon customers. In this setting, items with new pricing mechanisms may influence multiple customers who view them, while each customer may encounter multiple items subject to different pricing strategies.
\end{example}

% \begin{example}[Cluster randomization] 
% \end{example}

% \begin{example}[Geographical clustering]
%     \cite{rolnick2019randomized}
% \end{example}

% \begin{example}[Teacher-student]
%     A teacher can teach many students, and a student can take multiple classes. 
% \end{example}

% \begin{example}[Airbnb two-sided market]
%     \cite{li2022interference, fradkin2017search}
%     \cite{li2022interference} focuses on two-sided marketplace with $N$ listings and $M$ customers and studies two online experiments: customer-side randomization (CR) and listing-side randomization (LR). In CR experiments, 
% \end{example}

% \begin{example}[COVID and shelter-in-place]
% \end{example}

% \begin{example}[Causal inference with dyadic outcomes]
% \cite{d2016causal}    
% \end{example}

\subsection{Potential outcomes and causal estimand in bipartite experiments}
Consider a finite population with $m$ intervention units and $n$ outcome units. In bipartite experiments, the treatment is randomly assigned at the intervention unit level, while the outcomes of interest are measured at the outcome unit level. On the intervention unit side, the $m$ units are randomly assigned to treatment or control arms. Let $Z_k$ denote the binary treatment status of intervention unit $k$, where $Z_k$ equals $1$ if intervention unit $k$ is assigned to treatment and $0$ otherwise. Each outcome unit $i$ has $2^m$ potential outcomes $Y_i(\bz)$, where $\bz=(z_1,\ldots,z_m)\in\{0,1\}^{m}$ with $z_k=0,1$ for $k=1,\ldots,m$. The observed outcome is $Y=Y(\bZ)$, where $\bZ = (Z_1,\ldots,Z_m)\in\{0,1\}^{m}$ represents the assigned treatment vector. 

We focus on the total treatment effect
\begin{equation*}
    \tau = n^{-1}\sumi \left\{ \Yt{i} - \Yc{i} \right\}, 
\end{equation*}
where $\bOne$ and $\bZero$ represent the all-treated and all-control treatment assignment vectors, respectively. The parameter $\tau$ captures the difference in the average potential outcomes when all intervention units are treated versus when none are treated. It is of policy interest because it quantifies the overall impact of an intervention when applied universally across all intervention units. In many real-world applications, policymakers are concerned with understanding the aggregate consequences of fully implementing a treatment. It is a widely studied estimand in settings with interference, such as spatial experiments \citep{yu2022estimating, harshaw2023design} and temporal experiments \citep{liang2025randomization}. Define $\mu_1=n^{-1}\sumi\Yt{i}$ and $\mu_0=n^{-1}\sumi\Yc{i}$, so that $\tau=\mu_1-\mu_0$.

\subsection{Features of the bipartite graph and the bipartite interference assumption}

In bipartite experiments, intervention units and outcome units are connected through a bipartite graph, represented by a known, fixed $m\times n$ adjacency matrix $\bG$. The $(k,i)$-th entry $\G_{ki}$ equals $1$ if outcome unit $i$ is connected to intervention unit $k$, and $0$ otherwise, for $k=1,\ldots,m$ and $i=1,\ldots,n$. 

Define $\cG_{+i} \subset \{1,\ldots,m\}$ as the subset of indices of intervention units connected to outcome unit $i$, i.e., $k\in \cG_{+i}$ if and only if $\G_{ki}=1$. Let $\G_{+i} = |\cG_{+i}| = \sumk \G_{ki}$ denote the number of intervention units connected to $i$, thus $\max_{1\leq i\leq n}\G_{+i}$ is the maximum number of connections any outcome unit has. Define $\cG_{k+} \subset \{1,\ldots,n\}$ as the subset of indices of outcome units connected to intervention unit $k$, i.e., $i\in\cG_{k+}$ if and only if $G_{ki}=1$. Let $\G_{k+} = |\cG_{k+}| = \sumi G_{ki}$ denote the number of outcome units connected to intervention unit $k$, thus $\max_{1\leq k\leq m}\G_{k+}$ is the maximum number of outcome units any intervention unit is connected to.

The general potential outcome notation $Y_i(\bz)$ allows for arbitrary interference, making it overly general and intractable for practical analysis. To impose structure on the interference pattern, we introduce the following bipartite interference assumption which leverages the bipartite graph structure.

\begin{assumption}
\label{assump::exposure}
    $Y_i(\bz) = Y_i(\bz_{\cG_{+i}})$, where $\bz_{\cG_{+i}}$ denotes the subvector of $\bz$ corresponding to the intervention units in $\cG_{+i}$.
\end{assumption}

Assumption~\ref{assump::exposure} requires that the potential outcomes of outcome unit $i$ depend only on the treatment status of the intervention units it connects to. Unlike the classic setting, the potential outcome $Y_i(\bz)$ depends on the subvector $\bz_{\cG_{+i}}$, whose dimension varies across outcome units according to the bipartite graph $\G$. Under Assumption~\ref{assump::exposure}, the total treatment effect becomes $\tau=n^{-1} \sumi \{Y_i(\boldsymbol{1}_{G_{+i}})-Y_i(\boldsymbol{0}_{G_{+i}})\}$, with $\mu_1=n^{-1} \sumi Y_i(\boldsymbol{1}_{G_{+i}})$ and $\mu_0=n^{-1} \sumi Y_i(\boldsymbol{0}_{G_{+i}})$.

\subsection{Treatment assignment in bipartite experiments}
We assume that the treatment assignment follows Bernoulli randomization at the intervention unit level, formally stated as follows:
\begin{assumption}\label{assump::BR}
    $Z_i$,\ldots,$Z_m$ are independently and identically distributed as $\textup{Bernoulli}(p)$.
\end{assumption}

We focus on the Bernoulli randomization regime in Assumption~\ref{assump::BR}, where all intervention units share the same treatment probability $p$. A natural extension is to allow $Z_k$ to be independently assigned according to $\textup{Bernoulli}(p_k)$, allowing treatment probabilities $p_k$ to vary across intervention units, as in stratified randomization. Another possible extension is to observational studies under an ignorability assumption \citep{rosenbaum1983central}. Although these extensions are conceptually straightforward, they require very different theoretical analyses. We leave them to future research. 

\section{Estimation and inference}\label{sec::identification-estimation}
\subsection{Point estimator}\label{sec::point-estimator}
We first propose the H\'{a}jek estimator for $\tau$ in bipartite experiments. Let $T_i=  \prod_{k\in \mathcal{G}_{+i}} Z_k$ and $C_i= \prod_{k\in \mathcal{G}_{+i}} (1-Z_k)$
% $T_i=\indicator\{\sumk \G_{ki}(1-Z_k)=0\}$ and $C_i=\indicator\{\sumk \G_{ki}Z_k=0\}$ 
denote the indicators that all intervention units connected to outcome unit $i$ were assigned to the treatment and control arms, respectively. Under Assumption \ref{assump::BR}, we have $E(T_i) = p^{\G_{+i}}$ and $E(C_i) = (1-p)^{\G_{+i}}$. A natural Horvitz--Thompson-type estimator $n^{-1} \sumi T_i Y_i / p^{\G_{+i}} - n^{-1}\sumi C_i Y_i / (1-p)^{\G_{+i}}$ is unbiased for $\tau$. However, it is often unstable in finite samples. So we focus on the following H\'{a}jek-type weighting estimator $\hat\tau = \hat\mu_1 - \hat\mu_0$, where
\begin{eqnarray*}
    \hat\mu_1 &=& n^{-1} \sumi \frac{T_i Y_i}{p^{\G_{+i}}} \Big/ n^{-1}\sumi 
    \frac{T_i}{p^{\G_{+i}}}, \\ 
    \hat\mu_0 &=& n^{-1}\sumi \frac{C_i Y_i}{(1-p)^{\G_{+i}}} \Big/ n^{-1}\sumi \frac{C_i}{(1-p)^{\G_{+i}}}.
\end{eqnarray*}
The H\'{a}jek estimator is not unbiased, but in the next subsection, we show that it is consistent as $m\rightarrow\infty$. Since inference is driven by the treatment assignments $Z_k$'s on the $m$ intervention units, throughout the paper, we adopt the following asymptotic regime. Specifically, we consider a sequence of finite populations with the number of intervention units $m\rightarrow\infty$, the number of outcome units $n$ grows with $m$, and the bipartite graph expands accordingly, while maintaining the bounded covariates and outcome condition, as well as sparsity conditions specified in later sections. Otherwise, in the regime with a finite $m$, we should instead use finite-sample valid methods such as the Fisher randomization test \citep{ding2024first}. 

Intuitively, we need a sufficient number of outcome units with $T_i=1$ and $C_i=1$ to achieve the desired asymptotic properties of $\hat\mu_1$ and $\hat\mu_0$, which in turn depends on the sparsity of the bipartite graph. We provide formal characterizations in the following Sections~\ref{sec::consistency} and~\ref{sec::asym_normal}.

\subsection{Consistency of $\hat\tau$}
\label{sec::consistency}
In this subsection, we establish the consistency of $\hat\tau$. To do so, we impose the following regularity conditions.

\begin{assumption}
\label{assump::Y-X-max}
The potential outcomes and the covariates are bounded. 
\end{assumption}

Assumption~\ref{assump::Y-X-max} imposes boundedness on the potential outcomes and covariates, which enables the proof of the limiting theorems. While this assumption can be relaxed to some moment conditions, we keep its current form to simplify the presentation.

\begin{assumption}
\label{assump::Dmax}
$\Smax = O(1)$ and $\Dmax = o(n)$.
\end{assumption}

Assumption~\ref{assump::Dmax} restricts the density of the bipartite graph as $m$ increases. The condition $\Smax = O(1)$ requires that the maximum number of intervention units connected to any outcome unit is bounded by a constant. Intuitively, it requires no ``super influencer'' in the bipartite graph. Meanwhile, the condition $\Dmax = o(n)$ allows the number of outcome units connected to each intervention unit to diverge, but at a slower rate than $n$. Intuitively, it requires no ``super receiver'' in the bipartite graph. For instance, in Example~\ref{eg::power-plant}, $\Smax = O(1)$ requires that each neighborhood is affected by only a finite number of power plants, and $\Dmax = o(n)$ requires that no single power plant affects a dominant number of neighborhoods. These conditions are reasonable in Example~\ref{eg::power-plant} when neighborhoods are only affected by nearby power plants. In the special case of cluster randomization, the ratio $\Dmax/n$ reduces to the maximum of the relative cluster size in \cite{su2021model}. While Assumption~\ref{assump::Dmax} is reasonable in certain applications, such as Example~\ref{eg::power-plant}, it may be violated in general, in which case we need to target other causal estimands than $\tau$ or impose additional structural assumptions to aid in identification and estimation.

We now state the consistency result for $\hat\tau$.
\begin{theorem}[Consistency of $\hat\tau$]
\label{thm::consistency}
    Under Assumptions~\ref{assump::exposure}--\ref{assump::Dmax}, $\hat\tau$ converges in probability to $\tau$. 
\end{theorem}

In the supplementary material, we prove Theorem~\ref{thm::consistency} by showing $\hat\mu_z$ is consistent to $\mu_z$ for $z=0,1$. Specifically, $\hat\mu_1-\mu_1$ is asymptotically equivalent to $n^{-1}\sumi T_i(Y_i-\mu_1)/p^{\G_{+i}}$, the Horvitz--Thompson estimator based on the centered outcomes. We show that its variance converges to $0$, implying the consistency of $\hat\mu_1$ to $\mu_1$ by Chebyshev's inequality. A similar argument applies to $\hat\mu_0$, which consequently establishes the consistency of $\hat\tau$.

\subsection{Asymptotic normality}
\label{sec::asym_normal}

In this subsection, we establish the asymptotic normality of the point estimator $\hat\tau$. To do so, we further impose the following condition on the density of the bipartite graph.
\begin{assumption}
\label{assump::sparse}
Define two intervention units, $k_1$ and $k_2$, as connected if there exists at least one outcome unit linked to both. We assume that for any intervention unit $k$, the total number of intervention units connected to $k$ is bounded by an absolute constant $\B$:
$$\sum_{\ell\in[m] \backslash \{k\}}\indicator\{k,\ell\textup{ are connected} \} \leq \B, \quad (k = 1,\dots, m).$$
\end{assumption} 

Assumption~\ref{assump::sparse} imposes a sparsity condition on the degree of the bipartite graph. Assuming a bounded degree simplifies the presentation of the theoretical results, although our proof in the supplementary material accommodates cases where $B$ grows in some polynomial order of $n$. This sparsity condition can be justified in many bipartite experiments. For instance, in Example~\ref{eg::power-plant}, two power plants are defined as connected only if at least one neighborhood falls within a certain distance of both. If two power plants are geographically distant, they are not connected. Therefore, such a geographical network formation naturally restricts the sparsity of the network degrees. However, there are examples where Assumption~\ref{assump::sparse} is less likely to hold. In Example~\ref{eg::amazon}, items are connected in a dense pattern because each customer can browse a wide range of products, and browsing lists across different customers often overlap greatly. Again, in such cases, we need different estimation strategies and theoretical tools to analyze bipartite experiments with dense network structures.

We introduce additional notation for the potential outcomes. Let $\tilde{Y}_i(\bz)=Y_i(\bz)-n^{-1}\sumi Y_i(\bz)$ denote the centered potential outcomes for outcome unit $i$, and let $\tilde{\bY}(\bz)=(\tilde{Y}_1(\bz), \ldots, \tilde{Y}_n(\bz))^{\T}$ denote the vector of all centered potential outcomes under treatment assignment $\bz$. Define the following matrices for $i,j=1,\ldots,n$:
\begin{eqnarray}
    (\Lambda_1)_{i,j} = p^{-|\cG_{+i} \cap \cG_{+j}|} - 1, \ 
    (\Lambda_0)_{i,j} = (1-p)^{-|\cG_{+i} \cap \cG_{+j}|} - 1, \  
    (\Lambda_{\tau})_{i,j} = \indicator\{\cG_{+i} \cap \cG_{+j} \neq \varnothing\}. \label{eqn::lambda_matrics}
\end{eqnarray} 
Define
\begin{eqnarray}
    v_n &=& n^{-2}\left\{ \tilde\bY(\bOne)^{\T} \Lambda_1\tilde\bY(\bOne) + \tilde\bY(\bZero)^{\T} \Lambda_0
\tilde\bY(\bZero) +  
2 \tilde\bY(\bOne)^{\T} \Lambda_{\tau} \tilde\bY(\bZero) \right\}. \label{eqn::asym_var}
\end{eqnarray}

We have the following theorem on the asymptotic normality of $\hat\tau$.

\begin{theorem}[Asymptotic normality of $\hat\tau$]
\label{thm::asym_dist}
Under Assumptions~\ref{assump::exposure}--\ref{assump::sparse}, if $v_n$ is non-degenerate in the sense that $m^{-1/2}(\Dmax/n)^{-2}v_n\rightarrow \infty$, we have
\begin{eqnarray*}
v_n^{-1/2}(\hat\tau - \tau) & \rightarrow & \cN(0, 1)
\end{eqnarray*} 
in distribution.
\end{theorem}

In Section~\ref{sec::app_clt} of the supplementary material, we provide a general central limit theorem that is useful for the derivation of the asymptotic distribution of the estimators under Bernoulli randomization, which is also of independent interest for studying the asymptotic distribution of a class of statistics. The proof relies on constructing a martingale difference sequence from the estimator and applying the martingale central limit theorem of \cite{hall2014martingale}.

The asymptotic variance in~\eqref{eqn::asym_var} involves three $n\times n$ matrices $\Lambda_1$, $\Lambda_0$, and $\Lambda_{\tau}$, all of which rely on the structure of the bipartite graph. Specifically, they depend on $|\cG_{+i}\cap \cG_{+j}|$, the number of intervention units shared by any pair of outcome units $i$ and $j$, which relates to the second-order inclusion probabilities discussed in \cite{mukerjee2018using}. 

The regularity condition $m^{-1/2}(\Dmax/n)^{-2}v_n\rightarrow \infty$ in Theorem~\ref{thm::asym_dist} imposes two key requirements: $\max_{1\leq k\leq m}G_{k+}$ not to diverge too fast to $\infty$ and the variance $v_n$ not to converge too fast to $0$. The rate condition is motivated by our central limit theorem established in Theorem~\ref{thm::Gamma-CLT} in the supplementary material. Intuitively, the non-degeneracy condition requires the potential outcomes to have a non-degenerate covariance structure. 

We further illustrate the variance formula $v_n$ and the regularity condition in Theorem~\ref{thm::asym_dist} using the classic Bernoulli randomization and cluster randomization as examples. 

\begin{example}[Bernoulli randomization over units]
\label{exmp::cre}
In classic Bernoulli randomization where the randomization units are identical to the outcome units, 
$$
\cG_{+i} \cap \cG_{+j}= 
\begin{cases}
1, & \text { if } i=j, \\ 
0, & \text { if } i\neq j.
\end{cases}
$$
Thus $\Lambda_1, \Lambda_0, \Lambda_{\tau}$ are all diagonal matrices and the asymptotic variance in~\eqref{eqn::asym_var} reduces to
\begin{eqnarray*}
    v_n &=& n^{-2}p(1-p) \sumi \left\{ \frac{\tilde Y_i(1)}{p} - \frac{\tilde Y_i(0)}{1-p}\right\}^2,
\end{eqnarray*}
which recovers the classic results in \citet[][Theorem 1]{miratrix2013adjusting}. Moreover, the formula of $v_n$ is also identical to the classic results by \cite{neyman1923application} under complete randomization, which is a slightly different treatment regime from Bernoulli randomization but has equivalent asymptotic variance \citep{hajek1960limiting}.

In this setting, $\Dmax = 1$ and $n = m$, the regularity condition requires $v_n$ to have a larger order than $n^{-3/2}$. This is automatically satisfied according to the standard results in the literature \citep[e.g.,][]{li2017general}, which typically gives the rate of $v_n = O(n^{-1})$ under mild assumptions on the potential outcomes. 
\end{example}

\begin{example}[Bernoulli randomization over clusters]
\label{exmp::cluster}
In a cluster randomization setting with $m$ clusters and the treatment assignment indicators $Z_k$'s are independent and identically distributed as $\textup{Bern}(p)$, we have
$$
\cG_{+i} \cap \cG_{+j}= \begin{cases}1, & \text { if } i,j \text{ belong to the same intervention unit}, \\ 0, & \text { otherwise}.\end{cases}
$$
If we order the outcome units according to the clusters they belong to, the matrices in~\eqref{eqn::lambda_matrics} take the following block-diagonal forms:
\begin{equation}
    \Lambda_{\tau} \ =\  
    \begin{pmatrix}
    1_{n_1\times n_1} & 0 & \cdots & 0 \\
    0 & 1_{n_2\times n_2} & \cdots & 0 \\
    \vdots & \vdots & & \vdots \\
    0 & 0 & \cdots & 1_{n_m\times n_m}
    \end{pmatrix}, \quad 
    \Lambda_1 \ =\ \frac{1-p}{p}\Lambda_{\tau}, \quad 
    \Lambda_0 \ =\ \frac{p}{1-p}\Lambda_{\tau},
    \label{eqn::lambdas_cluster}
\end{equation}
where $n_k$ is the total number of outcome units in cluster $k$ for $k=1,\ldots, m$. Therefore, the asymptotic variance in~\eqref{eqn::asym_var} simplifies to
\begin{eqnarray*}
    v_n &=& n^{-2}p(1-p)\sumk \left[  \sum_{i\in\cG_{k+}} \left\{\frac{\tilde Y_i(1)}{p} - \frac{\tilde Y_i(0)}{1-p}\right\} \right]^{2},
\end{eqnarray*}
which aligns with the result in Theorem 1 of \cite{su2021model}. While \cite{su2021model} considered complete randomization at the cluster level, these two assignment regimes result in asymptotically equivalent variances, similar to the reasons discussed in Example~\ref{exmp::cre}.

In this setting, the regularity condition requires $m^{3/2}(\max_{1\leq k\leq m}\G_{k+}/\bar{\G}_{\cdot +})^{-2} v_n \rightarrow \infty$, where $\bar{\G}_{\cdot +}=m^{-1}\sumk \G_{k+}$ is the average cluster size. If the clusters are balanced in size, it reduces to $m^{3/2}v_n\rightarrow\infty$.
\end{example}

\begin{remark}
\label{rmk::regularity_condition_super_population}
    We use a super population model to aid the interpretation of the regularity condition $m^{-1/2}(\Dmax/n)^{-2}v_n\rightarrow \infty$. Consider a special case where $\Dmax$ and $\Smax$ are both finite, and thus $m$ and $n$ are of the same order. The potential outcomes $(Y_i(\bOne), Y_i(\bZero))$ are generated independently from a bivariate normal distribution $\cN((0,0)^\T, \textup{diag}\{\sigma_1^2, \sigma_0^2\})$. We show in the supplementary material that 
    \begin{eqnarray*}
        n v_n - \left[n^{-1}\sigma_1^2\sum_{i=1}^n (p^{-\G_{+i}} - 1) + n^{-1}\sigma_0^2\sum_{i=1}^n \{(1-p)^{-\G_{+i}} - 1\}\right] \rightarrow 0
    \end{eqnarray*} 
    almost surely. 
    This suggests that $v_n \asymp n^{-1}$. Thus, the regularity condition is satisfied because with a bounded $\Dmax$, we have
    \begin{eqnarray*}
        m^{-1/2}\left(\Dmax/n\right)^{-2}v_n
        \asymp m^{1/2}
        \rightarrow \infty. 
    \end{eqnarray*}
\end{remark}

\subsection{Variance estimation}\label{sec::variance-estimation}
To conduct Wald-type inference based on the central limit theorem in Theorem~\ref{thm::asym_dist}, we need to estimate the asymptotic variance $v_n$. Similar to other design-based inference settings \citep{neyman1923application,lin2013agnostic,ding2024first}, $v_n$ is not estimable because its last term depends on the joint values of the two potential outcomes $\tilde{\bY}(\bOne)$ and $\tilde{\bY}(\bZero)$. Instead, we construct an upper bound, $v_{n,\textup{UB}}$, on $v_n$, which is estimable. Applying the Cauchy--Schwarz inequality $\var(\hat\mu_1 - \hat\mu_0) \leq  \{\var(\hat\mu_1)^{1/2}+\var(\hat\mu_0)^{1/2}\}^2$, we derive an upper bound for $v_n$, given by $v_{n,\textup{UB}}=(v_1^{1/2}+v_0^{1/2})^2$, where
\begin{eqnarray*}
    v_1 = n^{-2}\tilde\bY(\bOne)^{\T} \Lambda_1 \tilde\bY(\bOne), \quad v_0 = n^{-2}\tilde\bY(\bZero)^{\T} \Lambda_0 \tilde\bY(\bZero).
\end{eqnarray*}
We formally establish the validity of this proposed upper bound later in Theorem~\ref{thm::var_est}(a) by showing $v_{n,\textup{UB}}\geq v_n$.

Our proposed variance estimator is $\hat v_{n,\textup{UB}} = (\hat v_1^{1/2} + \hat v_0^{1/2})^2$,
where
\begin{equation*}
    \hat v_1 = n^{-2}\sum_{i,j} \frac{T_iT_j(Y_i - \hat\mu_1)(Y_j - \hat\mu_1)(\Lambda_1)_{i,j}}{p^{|\cG_{+i} \cup \cG_{+j}|}}, \quad \hat v_0 = n^{-2}\sum_{i,j} \frac{C_iC_j(Y_i - \hat\mu_0)(Y_j - \hat\mu_0)(\Lambda_0)_{i,j}}{(1-p)^{|\cG_{+i} \cup \cG_{+j}|}}.
\end{equation*}
The terms $\hat v_1$ and $\hat v_0$ are sample analogues of $v_1$ and $v_0$, respectively. Both terms involve double summations over all pairs of outcome units $i$ and $j$, with inverse probability weighting based on the expected joint treatment assignment probabilities $E(T_iT_j)=p^{|\cG_{+i} \cup \cG_{+j}|}$ and $E(C_iC_j)=(1-p)^{|\cG_{+i} \cup \cG_{+j}|}$. It aligns with standard methods in survey sampling and experimental design, where variance estimation incorporates second-order inclusion probabilities \citep{mukerjee2018using}.

The following Theorem~\ref{thm::var_est}(b) shows that $\hat v_{n,\textup{UB}}$ is a consistent estimator of the upper bound, ensuring that $\hat v_{n,\textup{UB}}$ provides a conservative estimate of the true asymptotic variance $v_n$. Therefore, we can construct a valid $(1-\alpha)$-level Wald-type large-sample confidence interval for $\hat\tau$ as $$[\hat\tau - q_{\alpha/2}\hat v_{n,\textup{UB}}^{1/2}, \hat\tau + q_{\alpha/2}\hat v_{n,\textup{UB}}^{1/2}],$$ where $q_{\alpha/2}$ denotes the upper $\alpha/2$ quantile of a standard normal distribution.

\begin{theorem}[Conservative variance estimator for $\hat\tau$]
\label{thm::var_est}
(a) The proposed upper bound satisfies $v_{n,\textup{UB}}\geq v_n$, with equality if and only if
    \begin{eqnarray}\label{eqn::consistent_var_est}
        \tilde\bY(\bOne)^{\T}\Lambda_{\tau}\tilde\bY(\bZero) &=& \{\tilde\bY(\bOne)^{\T}\Lambda_{1}\tilde\bY(\bOne)\}^{1/2} \{\tilde\bY(\bZero)^{\T}\Lambda_{0}\tilde\bY(\bZero)\}^{1/2}.
    \end{eqnarray}
    
(b) Assume Assumptions~\ref{assump::exposure}--\ref{assump::sparse}. The ratio $\hat v_{n,\textup{UB}}/v_{n,\textup{UB}}$ converges in probability to 1.

\end{theorem}

The key to deriving an upper bound of $v_n$ is to bound the unidentifiable term that depends on the joint values of two potential outcomes. In Bernoulli randomization in Example~\ref{exmp::cre}, $\Lambda_\tau=I_n$, and in cluster randomization in Example~\ref{exmp::cluster}, the matrix $\Lambda_\tau$ reduces to the block matrix in~\eqref{eqn::lambda_matrics}. In both cases, $\Lambda_\tau$ is positive semi-definite, which allows us to bound the $\tilde\bY(\bOne)^{\T}\Lambda_{\tau}\tilde\bY(\bZero)$ using the inequality $\tilde\bY(\bOne)^{\T}\Lambda_{\tau}\tilde\bY(\bZero)\leq \{\tilde\bY(\bOne)^{\T}\Lambda_{\tau}\tilde\bY(\bOne)\}^{1/2} \{\tilde\bY(\bZero)^{\T}\Lambda_{\tau}\tilde\bY(\bZero)\}^{1/2}$. However, in bipartite experiments with a general bipartite graph, $\Lambda_\tau$ is not guaranteed to be positive semi-definite, making the previous inequality invalid. Therefore, instead of bounding $\tilde\bY(\bOne)^{\T}\Lambda_{\tau}\tilde\bY(\bZero)$, we bound $v_n$ by applying the Cauchy--Schwarz inequality to the covariance term, $\cov(\hat\mu_1,\hat\mu_0)^2\leq\var(\hat\mu_1) \var(\hat\mu_0)$, with equality holding when Condition~\eqref{eqn::consistent_var_est} is satisfied. This condition formally characterizes when the variance estimator achieves consistency. It depends on both the values of the potential outcomes and the structure of the bipartite graph. To provide more intuition, we revisit Examples~\ref{exmp::cre} and~\ref{exmp::cluster} to illustrate the equality condition in the special cases.

\begin{continuance}{exmp::cre}
In the classic Bernoulli randomized experiment setting, condition~\eqref{eqn::consistent_var_est} reduces to 
\begin{equation*}
    \sumi \tilde Y_i(1) \tilde Y_i(0) = \left\{\sumi \tilde Y_i(1)^2 \sumi \tilde Y_i(0)^2\right\}^{1/2}, \label{eqn::consistent_var_est_cre}
\end{equation*}
which is equivalent to $\tilde Y_i(1) = \zeta_1 \tilde Y_i(0)$ for all $i=1,\ldots, n$ for some positive constant $\zeta_1>0$. Such type of condition from the Cauchy--Schwarz inequality also appeared in~\cite{imai2008variance} and Problem 4.5 of~\cite{ding2024first}. A special case satisfying $\tilde Y_i(1) = \zeta_1 \tilde Y_i(0)$ with $\zeta_1=1$ corresponds to the constant treatment effect scenario \citep{neyman1923application}, where $Y_i(1) - Y_i(0) = \tau$ for all outcome units $i=1,\ldots, n$. 
\end{continuance}

\begin{continuance}{exmp::cluster}
In the cluster experiment setting, condition~\eqref{eqn::consistent_var_est} reduces to 
$$\sumk \left\{\sum_{i\in \cG_{k+}} \tilde Y_i(1)\right\}\left\{\sum_{i\in\cG_{k+}}\tilde Y_i(0)\right\} = \left[\sumk \left\{\sum_{i\in\cG_{k+}}\tilde Y_i(1)\right\}^2 \sumk\left\{\sum_{i\in\cG_{k+}}\tilde Y_i(0)\right\}^2\right]^{1/2},$$ 
which is equivalent to $\sum_{i\in\cG_{k+}}\tilde Y_i(1) = \zeta_2 \sum_{i\in\cG_{k+}}\tilde Y_i(0)$ for all $k = 1,\ldots,m$ for some positive constant $\zeta_2>0$. \cite{aronow2017estimating} discussed a similar condition. A special case that satisfies this condition is when the cluster-specific average treatment effect on each cluster is a constant, i.e., $\G_{k+}^{-1}\{\sum_{i\in\cG_{k+}}Y_i(1) - \sum_{i\in\cG_{k+}}Y_i(0)\} = \tau$ for all $k=1,\ldots, m$. 
\end{continuance}

\section{Covariate adjustment in bipartite experiments}\label{sec::covariate}

\subsection{Optimal covariate adjustment: an oracle estimator among linearly adjusted estimators}\label{sec::cov-adj-method}
In this section, we propose a covariate adjustment strategy to improve efficiency in bipartite experiments. Covariate adjustment is a classic topic in randomized experiments. In the setting of completely randomized experiments, \citet{fisher1925statistical} proposed to use the analysis of covariance to improve estimation efficiency. \citet{freedman2008bregression} later reanalyzed the analysis of the covariance estimator and found that it does not guarantee efficiency improvement in completely randomized experiments. In response to this critique, \citet{lin2013agnostic} proposed an alternative regression-based covariate adjustment method that guarantees asymptotic efficiency gains. We generalize the method of \citet{lin2013agnostic} to obtain a covariate adjustment strategy in bipartite experiments. 

Consider the scenario where outcome-unit-level covariate information is available, denoted by $X_i$, which represents characteristics of the outcome units. In the context of bipartite experiments, the quantity $|\mathcal{S}_i|$, which is the number of intervention units connected to outcome unit $i$, naturally serves as an additional covariate derived from the bipartite graph structure. For simplicity, we center the covariates so that $n^{-1}\sumi X_i=0$. 
% Let $\tilde X_i$ denote the centered covariates, i.e., $\tilde X_i = X_i - n^{-1}\sumi X_i$.  
Consider a class of linearly adjusted estimators indexed by $(\beta_1,\beta_0)$: 
\begin{eqnarray*}
    \hat\tau(\beta_1,\beta_0) &=& n^{-1} \sumi \frac{T_i (Y_i-\beta_1^\T X_i)}{p^{\G_{+i}}} \Big/ n^{-1}\sumi 
   \frac{T_i}{p^{\G_{+i}}} - n^{-1}\sumi \frac{C_i (Y_i-\beta_0^\T X_i)}{(1-p)^{\G_{+i}}} \Big/ n^{-1}\sumi \frac{C_i}{(1-p)^{\G_{+i}}},
\end{eqnarray*}
where we replace $Y_i$ in the previous H\'{a}jek-type estimator $\hat\tau$ with linearly adjusted residuals. Further denote $\bX=(X_1, \ldots, X_n)^\T$ the centered covariate matrix including covariates of all $n$ outcome units. The covariate adjustment estimator $\hat\tau(\beta_1,\beta_0)$ has the following properties for any fixed $(\beta_1,\beta_0)$. Define the asymptotic variance function
\begin{eqnarray}
    v_n(\beta_1,\beta_0) &=& n^{-2}\left[ 
    \{\tilde\bY(\bOne) - \bX \beta_1\}^{\T} \Lambda_1\{\tilde\bY(\bOne) - \bX \beta_1 \} + \{\tilde\bY(\bZero) - \bX \beta_0\}^{\T} \Lambda_0\{\tilde\bY(\bZero) - \bX \beta_0 \}\right. \notag\\
    && \quad\ + \left. 
    2 \{\tilde\bY(\bOne) - \bX \beta_1\}^{\T} \Lambda_{\tau} 
    \{\tilde\bY(\bZero) - \bX \beta_0\} \right]. \label{eqn::asym_var_X}
\end{eqnarray}

\begin{proposition}[Consistency and asymptotic distribution of $\hat\tau(\beta_1,\beta_0)$]
\label{prop::asym_dist_cov_adj}
    Under Assumptions~\ref{assump::exposure}--\ref{assump::Dmax}, for any fixed $(\beta_1,\beta_0)$, $\hat\tau(\beta_1,\beta_0)$ converges in probability to $\tau$. Further, if Assumption~\ref{assump::sparse} holds and $v_n(\beta_1,\beta_0)$ is non-degenerate in the sense that $m^{-1/2}(\Dmax/n)^{-2}v_n(\beta_1,\beta_0)\rightarrow \infty$, we have
\begin{eqnarray*}
    v_n(\beta_1,\beta_0)^{-1/2}\left\{\hat\tau(\beta_1,\beta_0) - \tau\right\} \rightarrow \cN(0,1)
\end{eqnarray*}
in distribution.
\end{proposition}

Proposition~\ref{prop::asym_dist_cov_adj} states the analogous results to Theorem~\ref{thm::asym_dist} on the asymptotic distribution of the class of covariate-adjusted estimators. The results follow directly when we treat the linearly adjusted residuals of the potential outcomes, $Y_i(1) - \beta_1^{\T} X_i$ and $Y_i(0) - \beta_0^{\T} X_i$, as ``pseudo potential outcomes'' and apply Theorem~\ref{thm::asym_dist}. 

% The remaining step is to check that the conditions in Theorems~\ref{thm::consistency} and~\ref{thm::asym_dist} still hold with the pseudo potential outcomes. Assumptions~\ref{assump::exposure}, \ref{assump::Dmax} and~\ref{assump::sparse} still hold because the network structure remains the same. We finally check Assumption~\ref{assump::Y-X-max} on pseudo potential outcomes. By the assumption that potential outcomes and covariates are bounded, there exist constants $c_Y$ and $c_X$ such that $|Y_i(\bz)| \le c_Y$ and $|X_{ik}| \le c_X$. Therefore, the pseudo potential outcomes, $|Y_i(\bz) - \beta_{z}^{\T} X_i| \le c_Y + \|\beta_z\|_1 c_X$, are also bounded.
% \qed
Similar as in Section~\ref{sec::variance-estimation}, we can construct a conservative variance estimator for $\hat\tau(\beta_1,\beta_0)$. Denote the upper bound of $v_n(\beta_1,\beta_0)$ as $v_{n,\textup{UB}}(\beta_1,\beta_0)=\{v_1(\beta_1,\beta_0)^{1/2}+v_0(\beta_1,\beta_0)^{1/2}\}^2$, where
\begin{eqnarray*}
    v_1(\beta_1,\beta_0) &=& n^{-2} \{\tilde\bY(\bOne) - \bX\beta_1\}^{\T} \Lambda_1 \{\tilde\bY(\bOne) - \bX\beta_1\} \\
    v_0(\beta_1,\beta_0) &=& n^{-2}\{\tilde\bY(\bZero) - \bX\beta_0\}^{\T} \Lambda_0 \{\tilde\bY(\bZero) - \bX\beta_0\}.
\end{eqnarray*}
A consistent estimator of the upper bound is $\hat v_{n,\textup{UB}}(\beta_1, \beta_0)=\{\hat v_1(\beta_1,\beta_0)^{1/2}+\hat v_0(\beta_1,\beta_0)^{1/2}\}^2$, where
\begin{eqnarray*}
    \hat v_1(\beta_1,\beta_0) &=& n^{-2}\sum_{i,j} \frac{T_iT_j(Y_i - \hat\mu_1 - \beta_1^\T X_i)(Y_j - \hat\mu_1 - \beta_1^\T X_j)(\Lambda_1)_{i,j}}{p^{|\cG_{+i} \cup \cG_{+j}|}}, \\
    \hat v_0(\beta_1,\beta_0) &=& n^{-2}\sum_{i,j} \frac{C_iC_j(Y_i - \hat\mu_0 - \beta_0^\T X_i)(Y_j - \hat\mu_0 - \beta_0^\T X_j)(\Lambda_0)_{i,j}}{(1-p)^{|\cG_{+i} \cup \cG_{+j}|}}.
\end{eqnarray*}

To achieve optimal asymptotic efficiency with covariate-adjusted estimators, ideally, we want to minimize the asymptotic variance $v_n(\beta_1,\beta_0)$ in~\eqref{eqn::asym_var_X} over $(\beta_1,\beta_0)$. However, the third term in $v_n(\beta_1,\beta_0)$ is not estimable from observed data, as it depends on the joint values of the potential outcomes. Working with the formula of $v_n(\beta_1, \beta_0)$, it may seem infeasible to obtain the optimal $\beta_1$ and $\beta_0$ to minimize the variance. Fortunately, we find an equivalent formulation of the problem that leads to feasible covariate adjustment. Minimizing $v_n(\beta_1, \beta_0)$ is equivalent to maximizing the reduction in the asymptotic variance of the covariate-adjusted estimator $\hat\tau(\beta_1,\beta_0)$ relative to the unadjusted estimator $\hat\tau$. Define the efficiency gain function as $L(\beta_1, \beta_0) = v_{n}-v_{n}(\beta_1,\beta_0)$, which represents the difference in asymptotic variance between $\hat\tau$ and $\hat\tau(\beta_1,\beta_0)$. The following lemma provides a useful expression for $L(\beta_1,\beta_0)$.

\begin{lemma}
\label{lemma::L_beta1_beta0}
We can write the efficiency gain function as 
\begin{eqnarray*}
    L(\beta_1, \beta_0) &=& 
    -n^{-2}
    \begin{pmatrix}
    \beta_1 \\
    \beta_0
    \end{pmatrix}^{\T} 
    \begin{pmatrix}
    \bX^{\T} \Lambda_1 \bX & \bX^{\T} \Lambda_\tau \bX \\
    \bX^{\T} \Lambda_\tau \bX & \bX^{\T} \Lambda_0 \bX
    \end{pmatrix}
    \begin{pmatrix}
    \beta_1 \\
    \beta_0
    \end{pmatrix} + 
    2n^{-2}
    \begin{pmatrix}
    \bX^{\T}\Lambda_1 \tilde \bY(\bOne) + \bX^{\T}\Lambda_\tau \tilde \bY(\bZero) \\
    \bX^{\T}\Lambda_0 \tilde \bY(\bZero) + \bX^{\T}\Lambda_\tau \tilde \bY(\bOne)
    \end{pmatrix}^{\T}
    \begin{pmatrix}
    \beta_1 \\
    \beta_0
    \end{pmatrix}.
\end{eqnarray*}
\end{lemma}

Lemma~\ref{lemma::L_beta1_beta0} shows that $L(\beta_1, \beta_0)$ is estimable from the observed data since it does not involve the joint values of the potential outcomes. Thus, we maximize $L(\beta_1,\beta_0)$ to obtain the optimal covariate-adjusted estimator:
\begin{eqnarray}
    \max_{\beta_1,\beta_0}  L(\beta_1, \beta_0). \label{eqn::oracle-opt}
\end{eqnarray}

Since $L(\beta_1,\beta_0)$ is a quadratic function of $(\beta_1,\beta_0)$, the first-order condition for the optimization problem~\eqref{eqn::oracle-opt} leads to the system of equations
\begin{eqnarray*}
    \begin{pmatrix}
    \bX^{\T} \Lambda_1 \bX & \bX^{\T} \Lambda_\tau \bX \\
    \bX^{\T} \Lambda_\tau \bX & \bX^{\T} \Lambda_0 \bX
    \end{pmatrix}\begin{pmatrix}
    \beta_1 \\
    \beta_0
    \end{pmatrix} &=& 
    \begin{pmatrix}
    \bX^{\T}\Lambda_1 \tilde \bY(\bOne) + \bX^{\T}\Lambda_\tau \tilde \bY(\bZero) \\
    \bX^{\T}\Lambda_0 \tilde \bY(\bZero) + \bX^{\T}\Lambda_\tau \tilde \bY(\bOne)
    \end{pmatrix}.
\end{eqnarray*}

Define the matrix
\begin{eqnarray*}
    \Omega_{n,xx} &=&
    \begin{pmatrix}
    \bX^{\T} \Lambda_1 \bX & \bX^{\T} \Lambda_\tau \bX \\
    \bX^{\T} \Lambda_\tau \bX & \bX^{\T} \Lambda_0 \bX
    \end{pmatrix}.
\end{eqnarray*}
In Proposition~\ref{prop::psd} in the Supplementary Material, we show that $\Omega_{n,xx}$ is positive semi-definite, ensuring that solutions to the optimization problem~\eqref{eqn::oracle-opt} exist. When $\Omega_{n,xx}$ is full-rank, the solution to~\eqref{eqn::oracle-opt} is unique and has a closed-form expression. When $\Omega_{n,xx}$ is not full-rank, there could exist multiple solutions. The non-uniqueness of the optimal covariate-adjusted estimators has been previously documented by \cite{li2020rerandomization} in the complete randomization setting. The proposed Lin's estimator \citep{lin2013agnostic} is one of the multiple optimal solutions.

To handle cases where $\Omega_{n,xx}$ is not full-rank, we use the Moore--Penrose pseudoinverse of $\Omega_{n,xx}$ to obtain a particular solution, denoted as $(\tilde\beta_1,\tilde\beta_0)$:
\begin{eqnarray}
    \begin{pmatrix}
    \tilde\beta_1 \\
    \tilde\beta_0
    \end{pmatrix} &=& 
    \begin{pmatrix}
    \bX^{\T} \Lambda_1 \bX & \bX^{\T} \Lambda_\tau \bX \\
    \bX^{\T} \Lambda_\tau \bX & \bX^{\T} \Lambda_0 \bX
    \end{pmatrix}^{\dagger}
    \begin{pmatrix}
    \bX^{\T}\Lambda_1 \tilde \bY(\bOne) + \bX^{\T}\Lambda_\tau \tilde \bY(\bZero) \\
    \bX^{\T}\Lambda_0 \tilde \bY(\bZero) + \bX^{\T}\Lambda_\tau \tilde \bY(\bOne)
    \end{pmatrix}, \label{eqn::oracle_solution}
\end{eqnarray}
where $M^{\dagger}$ denotes the Moore--Penrose pseudoinverse of a matrix $M$. 

The formulation of~\eqref{eqn::oracle-opt} ensures an improvement in asymptotic variance because by definition of $(\tilde\beta_1,\tilde\beta_0)$, $L(\tilde\beta_1,\tilde\beta_0) \geq L(0,0) \geq 0$. Therefore, the covariate adjustment estimator $\hat\tau(\tilde\beta_1,\tilde\beta_0)$ has an asymptotic variance no larger that of $\hat\tau$, i.e., $v_n(\tilde\beta_1,\tilde\beta_0)\leq v_n$. 

The estimator $\hat\tau(\tilde\beta_1,\tilde\beta_0)$ is the optimal covariate-adjusted estimator at the population level. However, since $(\tilde{\beta}_1,\tilde\beta_0)$ depends on the unknown potential outcomes, it cannot be directly computed from the observed data. In the next subsection, we introduce a feasible covariate-adjusted estimator that can be implemented using observed data.

\subsection{Estimation and inference based on a feasible covariate adjustment estimator}
In this subsection, we construct an estimator for the population-level solution in~\eqref{eqn::oracle_solution}. Since $(\tilde\beta_1,\tilde\beta_0)$ depends on unobserved potential outcomes, we estimate it using a sample analogue based on inverse probability weighting, similar to the variance estimation in Section~\ref{sec::variance-estimation}. Specifically, we define the estimator
\begin{eqnarray*}
    \begin{pmatrix}
    \hat\beta_1 \\
    \hat\beta_0
    \end{pmatrix} &=&
    \begin{pmatrix}
    \bX^{\T} \Lambda_1 \bX & \bX^{\T} \Lambda_\tau \bX \\
    \bX^{\T} \Lambda_\tau \bX & \bX^{\T} \Lambda_0 \bX
    \end{pmatrix}^{\dagger}
    \begin{pmatrix}
    \sum_{i,j} \frac{T_iT_j X_i (Y_j - \hat\mu_1) (\Lambda_1)_{i,j}}{p^{|\cG_{+i} \cup \cG_{+j}|}} + \sum_{i,j} \frac{C_iC_j X_i (Y_j - \hat\mu_0) (\Lambda_\tau)_{i,j}}{(1-p)^{|\cG_{+i} \cup \cG_{+j}|}} \\
    \sum_{i,j} \frac{T_iT_j X_i (Y_j - \hat\mu_1) (\Lambda_\tau)_{i,j}}{p^{|\cG_{+i} \cup \cG_{+j}|}} + \sum_{i,j} \frac{C_iC_j X_i (Y_j - \hat\mu_0) (\Lambda_0)_{i,j}}{(1-p)^{|\cG_{+i} \cup \cG_{+j}|}}
    \end{pmatrix}.
\end{eqnarray*}

Here, we apply the same inverse probability weighting approach as in Section~\ref{sec::variance-estimation} to estimate $\bX^{\T}\Lambda_1 \tilde \bY(\bOne) + \bX^{\T}\Lambda_\tau \tilde \bY(\bZero)$ and $\bX^{\T}\Lambda_0 \tilde \bY(\bZero) + \bX^{\T}\Lambda_\tau \tilde \bY(\bOne)$. We can also use this approach to construct estimates of $\bX^{\T}\Lambda_{*}\bX$ for $*=1,0,\tau$. However, instead of the sample analogues, we directly use the population quantities. This strategy has been explored to reduce the finite-sample bias of covariate-adjusted estimators \citep{chang2024exact, lu2023debiased}. 

With the estimated coefficients, we propose to use the covariate-adjusted point estimator $\hat\tau(\hat\beta_1,\hat\beta_0)$. To perform inference, we estimate the variance of $\hat\tau(\hat\beta_1,\hat\beta_0)$ using a similar approach described in Section~\ref{sec::cov-adj-method}, replacing $(\beta_1,\beta_0)$ with the estimated coefficients $(\hat\beta_1,\hat\beta_0)$. The resulting variance estimator is then $\hat v_{n,\textup{UB}}(\hat\beta_1,\hat\beta_0)$. The corresponding $(1-\alpha)$-level Wald-type confidence interval is 
$$[\hat\tau(\hat\beta_1, \hat\beta_0) - q_{\alpha/2}\hat v_{n,\textup{UB}}(\hat\beta_1,\hat\beta_0)^{1/2}, \hat\tau(\hat\beta_1, \hat\beta_0) + q_{\alpha/2}\hat v_{n,\textup{UB}}(\hat\beta_1,\hat\beta_0)^{1/2}],$$ 
where $q_{\alpha/2}$ denotes the upper $\alpha/2$ quantile of the standard normal distribution.

We next establish the asymptotic properties of the proposed point estimator $\hat\tau(\hat\beta_1, \hat\beta_0)$ and the corresponding variance estimator $\hat v_{n,\textup{UB}}(\hat\beta_1,\hat\beta_0)$. To facilitate the discussion, we introduce the following assumption that imposes the existence of limiting values for several finite-population quantities. 
% Define $\Davg = n/m$ to be the average number of units in each group.  
\begin{assumption}
\label{assump::limit_values}
There exists a universal constant $\delta\ge 0$ such that, as $m\to\infty$, the following convergence holds:
\begin{eqnarray*}
    m^{-(1+\delta)}
    \begin{pmatrix}
        \tilde \bY(\bOne) &
        \bX
    \end{pmatrix}^{\T}
    \Lambda_1
    \begin{pmatrix}
        \tilde \bY(\bOne) &
        \bX
    \end{pmatrix}
    &\rightarrow&
    \begin{pmatrix}
        \Omega_{yy, 11} & \Omega_{yx, 11}\\
        \Omega_{yx,11}^{\T} & \Omega_{xx, 11}
    \end{pmatrix} \ =:\ \Omega_{11}, \\
    m^{-(1+\delta)}
    \begin{pmatrix}
        \tilde \bY(\bZero) &
        \bX
    \end{pmatrix}^{\T}
    \Lambda_0
    \begin{pmatrix}
        \tilde \bY(\bZero) &
        \bX
    \end{pmatrix}
    &\rightarrow&
    \begin{pmatrix}
        \Omega_{yy, 00} & \Omega_{yx, 00}\\
        \Omega_{yx, 00}^{\T} & \Omega_{xx, 00}
    \end{pmatrix} \ =:\ \Omega_{00}, \\
    m^{-(1+\delta)}
    \begin{pmatrix}
        \tilde \bY(\bOne) &
        \bX
    \end{pmatrix}^{\T}
    \Lambda_\tau
    \begin{pmatrix}
        \tilde \bY(\bZero) &
        \bX
    \end{pmatrix}
    &\rightarrow&
    \begin{pmatrix}
        \Omega_{yy, 10} & \Omega_{yx, 10}\\
        \Omega_{yx, 01}^{\T} & \Omega_{xx, 10}
    \end{pmatrix} \ =:\ \Omega_{10}, 
\end{eqnarray*}    
where $\Omega_{yy,zz}>0$ and $\Omega_{xx,zz}$ are positive definite matrices for $z=0,1$. 
\end{assumption}

Assumption~\ref{assump::limit_values} requires that the weighted covariance matrices 
of the potential outcomes and covariates converge to limiting values not depending on $n$ as $m\rightarrow \infty$. In the special case of complete randomized experiments without interference, it reduces to the assumption in Theorem 5 in \cite{li2017general}. The rate $m^{-(1+\delta)}$ serves as a scaling factor that stabilizes the weighted covariance matrices and generally depends on the structure of the bipartite graph. For example, consider the special case of cluster randomization in Example~\ref{exmp::cluster} where all clusters have the same size, i.e., $n_1 = \cdots = n_m = \bar{n} \asymp m^{\delta_0}$. Then, we can compute
\begin{align*}
    m^{-(1+\delta)}\tilde Y(\bOne)^\T \Lambda_1 \tilde Y(\bOne) = m^{-(1+\delta)}\sumk \left\{  \sum_{i\in\cG_{k+}}  \tilde Y_i(1) \right\}^{2}
    = \frac{m \bar{n}^2}{m^{1+\delta}} \cdot \frac{1}{m}\sumk \left\{  \frac{1}{\bar{n}}\sum_{i\in\cG_{k+}}  \tilde Y_i(1) \right\}^{2},
\end{align*}
which has a stable limit with $\delta = 2\delta_0$ if the potential outcomes have stable moments. In general, the rate $\delta$ depends on the degree pattern of the bipartite graph and plays a key role in determining the asymptotic variance of the estimator.

We summarize the asymptotic properties of the covariate-adjusted point estimator $\hat\tau(\hat\beta_1,\hat\beta_0)$ and the conservativeness of its variance estimator $\hat v_{n,\textup{UB}}(\hat\beta_1, \hat\beta_0)$ in the following theorem.

\begin{theorem}[Asymptotic properties of $\hat\tau(\hat\beta_1,\hat\beta_0)$ and variance estimato r $\hat v_{n,\textup{UB}}(\hat\beta_1, \hat\beta_0)$]
\label{thm::cov_adj_clt}
Assume Assumptions~\ref{assump::exposure}--\ref{assump::Dmax} and~\ref{assump::limit_values}.

(a) $v_n(\tilde\beta_1,\tilde\beta_0)^{-1/2}\{\hat\tau(\hat\beta_1,\hat\beta_0)-\hat\tau(\tilde\beta_1,\tilde\beta_0)\}=o_{P}(1)$.

(b) If $m^{1/2+\delta}/(\Dmax)^2\to\infty$, then $\hat\tau(\hat\beta_1,\hat\beta_0)$ converges in probability to $\tau$. Further, if Assumption~\ref{assump::sparse} holds, we have
\begin{eqnarray*}
    \left\{v_n(\tilde\beta_1,\tilde\beta_0)\right\}^{-1/2}\left\{\hat\tau(\hat\beta_1,\hat\beta_0) - \tau\right\} & \rightarrow & \cN(0, 1)
\end{eqnarray*}
in distribution. 

(c) If Assumption~\ref{assump::sparse} holds, then $\hat v_{n,\textup{UB}}(\hat\beta_1, \hat\beta_0)$ is a conservative variance estimator of the asymptotic variance. This follows from the facts that $\hat v_{n,\textup{UB}}(\hat\beta_1, \hat\beta_0) / v_{n,\textup{UB}}(\tilde\beta_1,\tilde\beta_0)$ converges in probability to 1 and that $v_{n,\textup{UB}}(\tilde\beta_1,\tilde\beta_0) \geq v_n(\tilde\beta_1,\tilde\beta_0)$.
\end{theorem}

Theorem~\ref{thm::cov_adj_clt}(a) establishes the asymptotic equivalence between the covariate adjustment estimator when plugging in a global optimum $(\tilde\beta_1,\tilde\beta_0)$ and its estimator $(\hat\beta_1,\hat\beta_0)$. Theorem~\ref{thm::cov_adj_clt}(b) provides the consistency and asymptotic normality of $\hat\tau(\hat\beta_1,\hat\beta_0)$, implying that the covariate-adjusted estimator can achieve variance reduction compared to the unadjusted estimator. Theorem~\ref{thm::cov_adj_clt}(c) proves the convergence and conservativeness of the proposed variance estimator $\hat v_{n,\textup{UB}}(\hat\beta_1, \hat\beta_0)$, ensuring valid inference.

\subsection{Final remarks on covariate adjustment}

We present two remarks to end this section.

\begin{remark}[On the formulation of the optimization problem]
    We focus on maximizing the asymptotic efficiency gain of the covariate-adjusted estimator relative to the unadjusted estimator $\hat\tau$. By the formulation in~\eqref{eqn::oracle-opt}, the proposed covariate adjustment strategy ensures a reduction in asymptotic variance. However, the estimated variance may become even larger after covariate adjustment due to the conservatives of the variance estimator. Despite the possibility, such an occurrence is unlikely under realistic data-generating processes. Indeed, we observe a reduction in both the asymptotic variance and the estimated variance in all settings of our simulation study in Section~\ref{sec::simulation}.

    An alternative approach would be to directly minimize the variance estimator. However, such a strategy provides no guarantee of reducing the asymptotic variance itself and does not ensure better asymptotic power. 
    
    A more structured approach is to impose constraints on the optimization problem. One option is to maximize the asymptotic variance reduction, $L(\beta_1,\beta_0)$, subject to the constraint that the estimated variance does not increase. Another option is to minimize the estimated variance subject to the constraint that the true asymptotic variance does not increase, i.e., $L(\beta_1,\beta_0)\geq 0$. Both constrained optimization problems are feasible because they involve convex objective functions and convex constraints. However, they are more computationally challenging. Given these considerations, we adopt the current optimization formulation as it is the simplest choice that provides theoretical guarantees.
\end{remark}

\begin{remark}[Regression-assisted analysis of the estimators] 
    The unadjusted estimator $\hat\tau$ is identical to the coefficient of $T_i$ in a weighted least squares regression of $Y_i$ on $T_i$ with weights $T_i/p^{\G_{+i}}+C_i/(1-p)^{\G_{+i}}$. A natural extension of the regression approach to incorporate covariates information is to include $X_i$ in the weighted least squares regression model, with or without interaction with $T_i$, as discussed in \cite{zhao2022reconciling} and \cite{gao2023causal}. However, while such regression-based estimators may appear intuitive, they do not provide theoretical guarantees for efficiency improvement. Moreover, the corresponding robust standard errors do not have design-based properties without strong assumptions. Nevertheless, developing regression-based estimators and robust standard errors that retain valid design-based properties remains an important area in analyzing bipartite experiments.
\end{remark}

% \begin{proposition}[Improvement in both asymptotic variance and conservative variance estimator]
% \label{prop::both_improve}
% The covariate adjustment gives the following improvement for inference: 
% \begin{enumerate}
%     \item The conservative variance estimator for $\hat\tau(\hat\beta_1,\hat\beta_0)$ is no larger than that without covariate adjustment, i.e., $\hat v_{n,\textup{UB}}(\hat\beta_1, \hat\beta_0)\leq \hat v$.
%     \item Moreover, $\hat\tau(\hat\beta_1,\hat\beta_0)$ has an asymptotic variance no larger than $\hat\tau$.
% \end{enumerate}
% \end{proposition}

\section{Numerical examples}
\label{sec::simulation}
\subsection{Simulated bipartite experiments}
In this subsection, we conduct simulation studies to evaluate the finite-sample performance of our proposed estimators. The simulation starts by generating a bipartite graph consisting of two types of nodes: intervention unit nodes and outcome unit nodes. 

% To generate the degree $\G_{+i}$ of each outcome unit node, we sample from a normal distribution $\mathcal{N}\{(\Smax+1)/2,(\Smax-1)/6\}$ and round the sampled value to the nearest integer. This choice of distribution ensures that most sampled degrees fall within the range $[1,\Smax]$, thereby minimizing the need for truncation. 
The degree of each outcome unit node is randomly chosen with equal probability from the range of 1 to the specified maximum degree. After determining the degree of each outcome unit, we randomly connect outcome unit $i$ to $\G_{+i}$ distinct intervention unit nodes, forming the bipartite graph. Once the bipartite graph is generated and adjusted, we keep it fixed for the remainder of the simulation study.

% To ensure that the generated bipartite graph satisfies the sparsity condition in Assumption~\ref{assump::sparse}, we introduce an additional adjustment step. For each degree $s$, we examine the number of connected intervention unit sets through individuals with degree $s$. If the count surpasses a predefined upper limit, we break a random subset of the connections. Specifically, we break links between outcome units and intervention units that are part of the same overly connected set. We then randomly establish new connections with other intervention unit sets that were not previously overly connected, ensuring that the total degree of each individual is unchanged. Once the bipartite graph is generated and adjusted, we keep it fixed for the remainder of the simulation study.

We consider three different regimes of the data-generating process. For each regime, we generate covariates $X_i=(X_{1i},X_{2i})\sim (\textup{Uniform}[0,10])^2$ and the potential outcomes $Y_i(\bOne)$ and $Y_i(\bZero)$ from the following conditional distributions summarized in Table~\ref{tab::dgp_regimes}, with $\gamma=(5,5)^{\T}$ and $\alpha_i\sim \textup{Uniform}[0, 12]$. The treatment indicator $Z_k$'s are independently and identically distributed as $\textup{Bernoulli}(p)$ with $p=0.5$. 

\begin{table}[h]
\centering
{\small
\caption{Three regimes of data-generating process}
\label{tab::dgp_regimes}
\centering
\begin{tabular}{ccc}
\hline
\hline
regime&  $Y_i(\bOne)$& $Y_i(\bZero)$\\
\hline 
R1 &  $\cN(5.5 + \gamma^\T X_i, 10)$& $ \cN(\gamma^\T X_i, 10)$\\
R2 &  $\cN(\alpha_i + \gamma^\T X_i, 15)$ & $\cN(\gamma^\T X_i, 15)$\\
R3&  $\cN(0.5\G_{+i} + 1.1\gamma^\T X_i, 15)$& $\cN(\gamma^\T X_i, 15)$\\
\hline
\end{tabular}
}
\end{table}

Table~\ref{tab::simulation_5000} reports the finite-sample performance of the two estimators $\hat\tau$ and $\hat\tau(\hat\beta_1,\hat\beta_0)$ with $n=5000$, $m=500$, and $\Smax=5$ based on 1000 Monte Carlo replications. In all three regimes, the two estimators both have small finite sample bias, and the proposed variance estimators are conservative, leading to valid inference with coverage rates larger than $95\%$. Compared with the naive estimator $\hat\tau$, the covariate-adjusted estimator has a smaller standard error and higher power under all regimes. Although our theory guarantees efficiency improvement only in asymptotic variance, in the numerical studies, we also observe smaller variance estimators and thus shorter constructed confidence intervals under all three regimes. 

\begin{table}[h]
{\small
\centering
\caption{Finite-sample performance of estimators $\hat\tau$ and $\hat\tau(\hat\beta_1,\hat\beta_0)$.}
\label{tab::simulation_5000}
\begin{tabular}{cccccccccccc}
\hline
\hline
&   \multicolumn{5}{c}{unadjusted estimator $\hat\tau$} &  & \multicolumn{5}{c}{covariate-adjusted estimator $\hat\tau(\hat\beta_1,\hat\beta_0)$} \\  
\cline{2-6} \cline{8-12}
regime & bias & $\textsc{se}$ &  $\hat{\textsc{se}}$ & \textsc{cr} & power & & bias & $\textsc{se}$ & $\hat{\textsc{se}}$ & \textsc{cr} & power\\
\hline
R1 & $-0.001$ & $1.485$ & $1.951$ & $0.984$ & $0.818$ & & $0.021$ & $0.836$ & $1.023$ & $0.969$ & $0.995$ \\
R2 & $0.090$  & $1.734$ & $2.197$ & $0.983$ & $0.851$ & & $-0.002$ & $1.256$ & $1.734$ & $0.964$ & $0.985$ \\
R3 & $-0.025$  & $1.686$ & $2.253$ & $0.985$ & $0.818$ & & $0.017$ & $1.165$ & $1.456$ & $0.971$ & $0.993$ \\
\hline
\end{tabular}
\vskip 1em  
{\small For each data-generating regime, we report the bias, standard error \textsc{se}, estimated standard deviation $\hat{\textsc{se}}$, coverage rate \textsc{cr} of the 95\% confidence interval constructed using the conservative variance estimator, and the power of the two estimators, $\hat\tau$ and $\hat\tau(\hat\beta_1,\hat\beta_0)$.}
}
\end{table}

\subsection{Simulation based on a real data example}
\label{sec::application}

In this subsection, we apply our estimators to analyze a real-world bipartite graph. We revisit the application in \cite{zigler2021bipartite} and \cite{papadogeorgou2019adjusting}. They study the causal effect of the installation of a selective non-catalytic reduction system at a power plant on the air pollution level in the nearby areas. The intervention is at the power plant level, i.e., each power plant is assigned to either the implementation of the new system or not. Since multiple power plants simultaneously influence a given area, and each power plant can potentially affect multiple areas, it forms a natural bipartite graph. To model the scenario, we simulate a bipartite randomized experiment based on the real-world bipartite structure between power plants and nearby areas. We take power plants as intervention units and air pollution monitors as outcome units. 

We construct our dataset using the power plant dataset from~\cite{papadogeorgou2019adjusting} and 2004 air pollution data at the monitor level from the United States Environmental Protection Agency's website. Additionally, we incorporate population information for the counties where the monitors are located. The initial dataset includes 95762 air quality monitors as outcome units and 473 coal or natural gas-burning power plants as intervention units.
% To address computational constraints, we randomly select 60\% of the monitors. 
% To prepare the dataset, we exclude outcome units with an arithmetic mean above the 90th percentile of all observations, those with an arithmetic mean close to zero, and those with a population size exceeding $10^6$. We calculate the distances between monitors and power plants using their longitude and latitude coordinates. 
The bipartite graph is constructed by linking each monitor to power plants located within a 15-kilometer radius. The maximum number of connections for each outcome unit is 2. Monitors and power plants with no connections are excluded from the analysis. The final dataset consists of 7871 outcome units and 273 intervention units.  %To emphasize the influence of the closest intervention units, we cap the number of power plants connected to each monitor at two.

We generate the potential outcomes from $Y_i(\bOne)= -1.5 + \mathcal{N}(\gamma_1^\T X_i,15)$ and $Y_i(\bZero) = \mathcal{N}(\gamma_0^\T X_i,15)$, where $\gamma_1 = (0.02, -1.75)^{\T}$, $\gamma_0 = (0.01, -1.5)^{\T}$. We consider two covariates of monitor $i$, denoted as $X_i = (X_{1i}, X_{2i})$. Specifically, $X_{1i}$ is the scaled population size of the county where monitor $i$ is located, and $X_{2i}$ is the distance between monitor $i$ and its nearest power plant. We conduct 1000 Monte Carlo replications. The treatments are independently and identically distributed as $\textup{Bern}(p)$ with $p=0.5$. 

Table~\ref{tab::real-world} reports the simulation results based on the real-world bipartite graph. The results indicate that both the unadjusted and covariate-adjusted estimators exhibit small biases in estimating the total treatment effect, and both strategies lead to valid but conservative confidence intervals. However, applying the covariate adjustment strategy introduced in Section~\ref{sec::covariate} leads to a reduction in both the standard error of the point estimator and the estimated variance. This reduction translates into a substantial improvement in statistical power, demonstrating the effectiveness of covariate adjustment in enhancing estimation efficiency.

% true effect: -3.328$
\begin{table}[h]
{\small
\centering
\caption{Simulation results based on real bipartite graph in the power plant application}
\label{tab::real-world}
\begin{tabular}{cccccc}
\hline
\hline
estimator & bias &$\textsc{se}$&  $\hat{\textsc{se}}$&  \textsc{cr} & power\\ 
\hline
$\hat\tau$ & $-0.015$ & $1.040$ & $1.162$ & $0.971$ & $0.838$\\ 
$\hat\tau(\hat\beta_1,\hat\beta_0)$ & $-0.032$ & $0.518$ & $0.771$ & $0.983$ & $0.987$\\
\hline
\end{tabular}
\vskip 1em
{\small We report the bias, standard error \textsc{se}, estimated standard error $\hat{\textsc{se}}$, coverage rate \textsc{cr} of the 95\% confidence interval constructed using the conservative variance estimator, and the power of the two estimators, $\hat\tau$ and $\hat\tau(\hat\beta_1,\hat\beta_0)$.}
}
\end{table}

\section{Discussion}\label{sec::discussion}
We propose a design-based causal inference framework for bipartite experiments. We generalize the classic stable unit treatment value assumption to the bipartite experiment setting and provide point and variance estimators for estimating the total treatment effect. These estimators are based on theoretical results that guarantee the consistency and asymptotic normality of the point estimator and the conservativeness of the variance estimator. We also propose covariate adjustment strategies that improve the efficiency of the point estimator.
This framework extends the design-based causal inference frameworks for completely randomized experiments and cluster randomized experiments.

There are several directions for further research. First, we focus on the total treatment effect, which compares all versus nothing treatment regimes. A natural extension is to consider more general causal parameters of interest. Second, our covariate adjustment strategy mainly focuses on the outcome-unit-level covariates. When intervention-unit-level covariates are also available, as an ad-hoc strategy, we can incorporate them by summarizing them at the outcome-unit level, for instance, by taking the average or sum of the covariate values across the intervention units connected to each outcome unit. However, a more rigorous and systematic method for incorporating intervention-unit-level covariates requires additional research. Third, a growing body of work has also explored experimental design under bipartite interference \citep{viviano2023causal, ugander2013graph, pouget2019variance, harshaw2023design}. We can explore optimal treatment assignment strategies and clustering methods to maximize the efficiency of the proposed estimator. We leave them for future research. 
% Finally, we only consider the case where randomization is at the intervention-unit level. Extensions to a two-sided randomization setting are of interest, motivated by online platform settings where randomization can happen simultaneously on both sides of the bipartite graph. The multiple randomization design proposed in \cite{masoero2024multiple} covers a two-sided randomization design with a fully connected bipartite. With interference induced by the bipartite graph, it extends the classic factorial experiment designs \citep{zhao2022reconciling}. Moreover, it allows for a richer structure of the outcomes, incorporating outcomes measured on both sides of the graph as well as dyadic outcomes along the edges. 

\bibliographystyle{apalike}
\bibliography{ref}

\newpage 
\appendix

\renewcommand{\thesection}{\Alph{section}}
\renewcommand{\thetheorem}{\Alph{section}.\arabic{theorem}}
\setcounter{theorem}{0}
\renewcommand{\thelemma}{\Alph{section}.\arabic{lemma}}
\setcounter{lemma}{0}
\renewcommand{\theproposition}{\Alph{section}.\arabic{proposition}}
\setcounter{proposition}{0}
\renewcommand{\thecorollary}{\Alph{section}.\arabic{corollary}}
\setcounter{corollary}{0}
\renewcommand{\thedefinition}{\Alph{section}.\arabic{definition}}
\setcounter{definition}{0}
\renewcommand{\thepage}{S\arabic{page}}
\setcounter{page}{1}
\renewcommand{\theequation}{\Alph{section}.\arabic{equation}}
\setcounter{equation}{0}
\renewcommand{\thetable}{S\arabic{table}}
\setcounter{table}{0}

\begin{center}
  \LARGE {\bf Supplementary Material}
\end{center}

Section~\ref{sec::app_clt} provides a general central limit theorem that is useful for analyzing bipartite experiments under Bernoulli randomization, which is also of independent interest for studying the asymptotic distribution of a class of statistics. 

Section~\ref{sec::proofs} provides two useful lemmas and proofs of all results in the main text. 

Section~\ref{sec::psd_lambda} provides properties of the matrices $\Lambda_1$, $\Lambda_0$, and $\Lambda_\tau$ appeared in the asymptotic variance $v_n$. 

We introduce additional notation for the supplementary material. Let $\plim(\cdot)$ denote the probability limit, $\avar(\cdot)$ denote the asymptotic variance, and $\acov(\cdot,\cdot)$ denote the asymptotic covariance. 

\section{A central limit theorem}
\label{sec::app_clt}

\subsection{A general statistic}
To prove the central limit theorem for $\hat\tau$, we first prove a central limit theorem for a general statistic. As a convention, we use $(k_1\ldots k_s)$ to denote an unordered $s$-tuple with $1\le k_1\neq\cdots\neq k_s \le m$. The statistic is called a random polynomial, which is defined as follows:
\begin{definition}[Random polynomial]\label{def::Gamma}
Let $\tilde Z_k$'s be independently and identically distributed (i.i.d.) copies of a random variable $\tilde Z$ with mean zero and variance $\sigma^2$. Let $\{a_{k_1\dots k_s}: k_1,\dots,k_s\in[m], ~ k_1\neq \dots \neq k_s\}$ be a set of $s$-dimensional arrays with $s=1,\dots,S$. Assume they are symmetric in their indices, i.e.,
\begin{eqnarray*}
    a_{k_1\dots k_s} &=& a_{k'_1\dots k'_s} \quad \text{if} \quad 
    \{k_1,\dots,k_s\} 
    = 
    \{k'_1,\dots,k'_s\}.
\end{eqnarray*}
A random polynomial is defined as follows:
\begin{eqnarray}
  \Gamma &=& \sum_{k_1\in[m]} a_{k_1}\tilde Z_{k_1}
  +
  \sum_{(k_1k_2) \subset [m]}a_{k_1k_2}\tilde Z_{k_1}\tilde Z_{k_2}+\dots+ \sum_{(k_1\dots k_S)\subset[m]} a_{k_1\dots k_S}\tilde Z_{k_1}\cdots \tilde Z_{k_S}. \label{eqn::Gamma}
\end{eqnarray}
\end{definition}
Definition \ref{def::Gamma} does not require $\tilde Z$ to be a centered Bernoulli variable, although our proofs for bipartite experiments do.  The random polynomial covers many special statistics. For example, many works have studied the quadratic forms of i.i.d. variables \citep{de1987central, wu2007limit, lu2023debiased}, which correspond to the case with $m = 2$. The $s$-th summation in \eqref{eqn::Gamma} is called a random symmetric polynomial and has been studied in the U-statistics literature \citep{korolyuk2013theory}. Our goal here is to provide a set of general conditions for the asymptotic normality of $\Gamma$, which will be used to quantify the asymptotic behavior of the proposed estimators in later sections.  

The mean and variance of \eqref{eqn::Gamma} is given by the following lemma:
\begin{lemma}[Mean and variance of the random polynomial]\label{lemma::mean-var-Gamma}
    For the random polynomial in~\eqref{eqn::Gamma}, we have $E(\Gamma) = 0$ and  
\begin{eqnarray*}
v_\Gamma \ =\  \var(\Gamma) &=& \sum_{k_1\in[m]} a_{k_1}^2\sigma^2
+
\sum_{(k_1k_2)\subset[m]} a_{k_1k_2}^2 \sigma^4
+
\dots 
+
\sum_{(k_1 \dots k_S) \subset [m]}a_{k_1\dots k_S}^2 \sigma^{2S}.
\end{eqnarray*}
\end{lemma}

Our main result in this section is the following central limit theorem for $\Gamma$.
\begin{theorem}
\label{thm::Gamma-CLT}
Assume that:
(a) $\tilde Z$ has a bounded fourth moment: $E{\tilde Z^4} \leq \nu^4_4$;

(b) The elements of the array $a$'s are bounded by some constant $\bar a_m$ that possibly depends on $m$;  

(c) There exists a universal constant $\B$ such that for all $k\in[m]$ and $s\in[S]$, 
    \begin{eqnarray*}
        \sum_{(k_1 \ldots k_s)\subset[m]\backslash \{k\}}\indicator\{|a_{kk_1\dots k_s}|\neq 0 \} &\leq& \B; 
    \end{eqnarray*}
    
(d) ${v_\Gamma}/{(m^{1/2} \bar{a}_m^2)} \rightarrow \infty$ as $m\rightarrow\infty$. 
    
Then $v_\Gamma^{-1/2}\Gamma \rightarrow \cN(0, 1)$ in distribution as $m\rightarrow\infty$.
\end{theorem}

We will use the martingale central limit theorem in \cite{hall2014martingale} to prove Theorem~\ref{thm::Gamma-CLT}. For completeness of our proof, we first review the martingale central limit theorem as the following Proposition~\ref{prop::martingale_clt}.
\begin{proposition}[Theorem 3.2 of \cite{hall2014martingale}]
\label{prop::martingale_clt}
Let $\left\{S_{n i}, \cF_{n i}, 1 \leq i \leq k_n, n \geq 1\right\}$ be a zero-mean, square-integrable martingale array with nested $\sigma$-fields: $\cF_{n, i} \subseteq \cF_{n+1, i}$, for $1 \leq  i \leq k_n$, $n \geq 1$. Denote the differences $\Delta_{n i} = S_{n i} - S_{n (i-1)}$ with $S_{n 0} = 0$. Suppose the following conditions hold: 

(a) Squared sum convergence: 
    \begin{eqnarray}
        \sum_{i=1}^{k_n} E(\Delta_{n i}^2\mid \cF_{n, i-1} ) &\rightarrow& \eta^2 \label{eqn::squared_sum_convergence}
    \end{eqnarray}
    in probability as $n\rightarrow\infty$ for an almost surely finite random variable $\eta^2$.

(b) Lindeberg condition: for all $\varepsilon>0$,
    \begin{eqnarray}
        \sum_{i=1}^{k_n} E(\Delta_{n i}^2 \indicator\left\{\left|\Delta_{n i}\right|>\varepsilon\right\} \mid \cF_{n, i-1}) &\rightarrow& 0 \label{eqn::lindeberg}
    \end{eqnarray}
    in probability as $n\rightarrow\infty$. 
    
    Then $S_{n k_n}=\sum_{i=1}^{k_n} \Delta_{n i}$ converges in distribution to some random variable with characteristics function $E\{\exp(-1/2 \eta^2 t^2 )\}$, where the expectation is taken over $\eta$.
\end{proposition}

As a standard result in probability theory, a sufficient condition for the Lindeberg condition~\eqref{eqn::lindeberg} is the following Lyapunov condition: for some $\delta > 0$,
\begin{eqnarray}
    \sum_{i=1}^{k_n} E\{|\Delta_{n i}|^{2+\delta}\} \rightarrow 0. \label{eqn::lyapunov}
\end{eqnarray}
In our proof below, we will prove the asymptotic normality of our proposed estimator by verifying \eqref{eqn::squared_sum_convergence} and \eqref{eqn::lyapunov}.

\subsection{Proof of Lemma \ref{lemma::mean-var-Gamma}}
$\Gamma$ has mean zero because each $\tilde Z_k$ has mean zero and is independent of each other. To compute the variance, we write 
\begin{eqnarray*}
    \Gamma = \sum_{s=1}^{S} \Gamma_s,\quad \text{with}\quad \Gamma_s =  
    \sum_{(k_1\dots k_s)\subset[m]} a_{k_1\dots k_s}\tilde Z_{k_1}\cdots \tilde Z_{k_s}.
\end{eqnarray*}
We start by showing that 
\begin{eqnarray*}
    \cov(\Gamma_s,\Gamma_r) = 0, \quad \text{for} \quad s < r. 
\end{eqnarray*}
By definition, 
\begin{eqnarray*}
    \cov(\Gamma_s,\Gamma_r) 
    = 
    E(\Gamma_s\Gamma_r)
    =
    \sum_{(k_1 \dots k_s)\subset[m]}\sum_{(l_1 \dots l_r)\subset[m]} a_{k_1\dots k_s}a_{l_1\dots l_r}
    E\{(\tilde Z_{k_1}\cdots \tilde Z_{k_s})
    (\tilde Z_{l_1}\cdots \tilde Z_{l_r})\}. 
\end{eqnarray*}
Because $s<r$, there is at least one $l\in\{l_1,\dots,l_r\}$ that is different from $\{k_1,\dots,k_s\}$. Therefore, by independence, we always have 
\begin{eqnarray*}
    E\{(\tilde Z_{k_1}\cdots \tilde Z_{k_s})
    (\tilde Z_{l_1}\cdots \tilde Z_{l_r})\} = 0,
\end{eqnarray*}
which implies that $\cov(\Gamma_s,\Gamma_r) = 0$. Then we have
\begin{eqnarray*}
    \var(\Gamma) = \sum_{s=1}^{S} \var(\Gamma_s). 
\end{eqnarray*}
Then we need to compute $\var(\Gamma_s)$:
\begin{eqnarray*}
    \var(\Gamma_s)
    =
    \sum_{(k_1 \dots k_s)\subset[m]}\sum_{(l_1 \dots l_s)\subset[m]} a_{k_1\dots k_s}a_{l_1\dots l_s}
    E\{(\tilde Z_{k_1}\cdots \tilde Z_{k_s})
    (\tilde Z_{l_1}\cdots \tilde Z_{l_s})\}. 
\end{eqnarray*}
Again, by independence, we always have 
$E\{(\tilde Z_{k_1}\cdots \tilde Z_{k_s})
(\tilde Z_{l_1}\cdots \tilde Z_{l_s})\} = 0$ if $(k_1\dots k_s) \neq (l_1\dots l_s)$. Hence, 
\begin{eqnarray*}
    \var(\Gamma_s)
    =
    \sum_{(k_1 \dots k_s)\subset[m]} a_{k_1\dots k_s}^2
    E\{(\tilde Z_{k_1}\cdots \tilde Z_{k_s})^2\}
    =
    \sum_{(k_1 \dots k_s)\subset[m]}a_{k_1\dots k_s}^2 \sigma^{2s}. 
\end{eqnarray*}
This concludes the proof. 
\qed

\subsection{Proof of Theorem~\ref{thm::Gamma-CLT}}
We prove Theorem~\ref{thm::Gamma-CLT} following three steps. We first construct a martingale difference sequence based on $\Gamma$. Next, we check the convergence of the summation of the conditional squared differences in~\eqref{eqn::squared_sum_convergence} with $\eta = 1$. Finally, we check the Lyapunov condition in~\eqref{eqn::lyapunov}, which ensures the Lindeberg condition in \eqref{eqn::lindeberg}. 

\noindent \textbf{Step I.} Let $\cF_{m,k}$ be the $\sigma$-algebra generated by $\tilde Z_1,\ldots, \tilde Z_k$, i.e., $\cF_{m,k} = \sigma\{\tilde Z_1, \ldots, \tilde Z_k\}$. For ease of notation, for any $k_1,\dots,k_{\ell}\in[m]$, define $\tilde Z_{k_1\ldots k_{\ell}}=\tilde Z_{k_1}\cdots\tilde Z_{k_{\ell}}$. Let 
\begin{eqnarray*}
    \Delta_{mk} &=& v_\Gamma^{-1/2}
    \sum_{s=1}^{ S\wedge k}\sum_{(k_1\ldots k_{s-1})\subset[k-1]} a_{k_1\ldots k_{s-1}k}\tilde Z_{k_1\dots k_{s-1}k},
\end{eqnarray*}
with $a_{\varnothing k}=a_k$ and $\tilde Z_{\varnothing}=1$. Then $\{\Delta_{mk}, \cF_{m,k}\}_{k=1}^m$ forms a martingale difference sequence and 
\begin{eqnarray*}
    \Gamma &=& \sum_{k=1}^m \Delta_{mk}.
\end{eqnarray*}    

\noindent \textbf{Step II.} Check the convergence of the summation of the conditional squared differences in~\eqref{eqn::squared_sum_convergence}. 

We show~\eqref{eqn::squared_sum_convergence} by computing the variance of its left-hand side:
\begin{eqnarray}
    && \var\left\{\sumk E(\Delta_{mk}^2\mid \cF_{m,k-1})\right\} \notag \\
    &=& \frac{\sigma^4}{v_\Gamma^2} \var\left( \sumk \sum_{s,r=1}^{S\wedge k}
    \sum_{\substack{(k_1\dots k_s)\subset[k-1] \\ (k_1'\dots k_r')\subset[k-1]
    }} a_{k_1\dots k_s k}a_{k_1'\dots k_r' k} \tilde Z_{k_1\dots k_s} \tilde Z_{k_1'\dots k_r'} \right) \notag \\
    &=& \frac{\sigma^4}{v_\Gamma^2} \left\{ \sum_{k,\ell=1}^{m} \sum_{s,r=1}^{S\wedge k}\sum_{t,u=1}^{S\wedge \ell}
    I(k,l,s,r,t,u) 
     \right\}, \label{eqn::var-sum-square}
\end{eqnarray}
where 
\begin{eqnarray*}
    I(k,l,s,r,t,u) = \sum_{\substack{
    (k_1\dots k_s)\subset[k-1] \\ 
    (k_1'\dots k_r')\subset[k-1]\\
    (\ell_1'\dots \ell_t')\subset[\ell-1]\\
    (\ell_1'\dots \ell_u')\subset[\ell-1]
    }}
    a_{k_1\dots k_s k}
    a_{k_1'\dots k_r' k} 
    a_{\ell_1\dots \ell_t \ell}
    a_{\ell_1'\dots \ell_u' \ell} \cov(\tilde Z_{k_1\dots k_s}\tilde Z_{k_1'\dots k_r'}, \tilde Z_{\ell_1\dots \ell_t}\tilde Z_{\ell_1'\dots \ell_u'}).
\end{eqnarray*}

Note that $\cov(\tilde Z_{k_1\dots k_s}\tilde Z_{k_1'\dots k_r'}, \tilde Z_{\ell_1\dots \ell_t}\tilde Z_{\ell_1'\dots \ell_u'}) \neq 0$ only if 
\begin{eqnarray*}
    \{(k_1\dots k_s)\cup(k'_1\dots k'_r)\} 
    \cap 
    \{(\ell_1\dots \ell_t)\cup(\ell'_1\dots \ell'_u)\} &\neq& \varnothing. 
\end{eqnarray*}
For the nonzero covariance, we have
\begin{eqnarray*}
    &&\left|\cov(\tilde Z_{k_1\dots k_s}\tilde Z_{k_1'\dots k_r'}, \tilde Z_{\ell_1\dots \ell_t}\tilde Z_{\ell_1'\dots \ell_u'})\right| \\
    &\leq & \left\{\var(\tilde Z_{k_1\dots k_s}\tilde Z_{k_1'\dots k_r'})\var(\tilde Z_{\ell_1\dots \ell_t}\tilde Z_{\ell_1'\dots \ell_u'})\right\}^{1/2} \\
    &\leq & \left\{E(Z^2_{k_1\dots k_s}Z^2_{k_1'\dots k_r'})E(Z^2_{\ell_1\dots \ell_t}Z^2_{\ell_1'\dots \ell_u'})\right\}^{1/2} \\
    &\leq & E(\tilde Z_{k_1\dots k_s}^4)^{1/4} E(\tilde Z_{k_1'\dots k_r'}^4)^{1/4} E(\tilde Z_{\ell_1\dots \ell_t}^4)^{1/4} E(\tilde Z_{\ell_1'\dots \ell_u'}^4)^{1/4}\\
    &\leq & \nu_4^{s+r+t+u} \leq \nu_4^{4(S-1)}. 
\end{eqnarray*}
Now, for a set $W\subset[m]$, we use $a_W$ to denote the coefficient in Definition \ref{def::Gamma} indexed by the elements in $W$. We can further bound~\eqref{eqn::var-sum-square} as
\begin{eqnarray*}
    &&\var\left\{\sumk E(\Delta_{mk}^2\mid \cF_{m,k-1})\right\} \\
    &\leq & \frac{\sigma^4 \nu_4^{4(S-1)} \bar{a}_m^4}{v_\Gamma^2} \left( \sum_{\substack{W_1,W_2,W_3,W_4\subset [m]: \\
    \max\{|W_1|, |W_2|, |W_3|, |W_4|\}\le S,\\
    W_1\cap W_2 \neq \varnothing,\\
    W_2\cap W_3 \neq \varnothing,\\
    W_3\cap W_4 \neq \varnothing
    }}
    \indicator\{|a_{W_1}|\neq 0\}
    \indicator\{|a_{W_2}|\neq 0\} 
    \indicator\{|a_{W_3}|\neq 0\}
    \indicator\{|a_{W_4}|\neq 0\}
    \right) \\
    &\leq & 
    \frac{\sigma^4 \nu_4^{4(S-1)} \bar{a}_m^4 (\B S^2)}{v_\Gamma^2} \left( \sum_{\substack{W_1,W_2,W_3\subset [m]: \\
    \max\{|W_1|, |W_2|, |W_3|\}\le S,\\
    W_1\cap W_2 \neq \varnothing,\\
    W_2\cap W_3 \neq \varnothing
    }}
    \indicator\{|a_{W_1}|\neq 0\}
    \indicator\{|a_{W_2}|\neq 0\} 
    \indicator\{|a_{W_3}|\neq 0\}
    \right) \\
    &\leq & 
    \frac{\sigma^4 \nu_4^{4(S-1)} \bar{a}_m^4 (\B^2 S^4)}{v_\Gamma^2} \left\{ \sum_{\substack{W_1,W_2\subset [m]: \\
    \max\{|W_1|, |W_2|\}\le S,\\
    W_1\cap W_2 \neq \varnothing
    }}
    \indicator\{|a_{W_1}|\neq 0\}
    \indicator\{|a_{W_2}|\neq 0\}
    \right\} \\
    &\leq & \frac{\sigma^4 \nu_4^{4(S-1)} \bar{a}_m^4 (\B^4 S^6) m }{v_\Gamma^2}\\
    &= & o(1),
\end{eqnarray*}
under the third assumed condition.

Also, we have
\begin{eqnarray*}
    E\left\{\sumk E(\Delta_{mk}^2\mid \cF_{m,k-1})\right\} &=& 1. 
\end{eqnarray*}
Therefore, by Chebyshev's inequality, 
\begin{eqnarray*}
    \sumk E(\Delta_{mk}^2\mid \cF_{m,k-1}) &\rightarrow& 1
\end{eqnarray*}
in probability.

\noindent\textbf{Step III.} Check the Lyapunov condition in~\eqref{eqn::lyapunov}. 

We have
\begin{eqnarray*}
    && \sumk E(\Delta_{mk}^4) \\
    &=& \frac{1}{v_\Gamma^2} 
    \left\{\sum_{s,r,t,u}^{{S}\wedge k} 
    \sum_{\substack{
    (k_1\dots k_s)\subset[k-1] \\ 
    (k_1'\dots k_r')\subset[k-1]\\
    (k_1''\dots k_t'')\subset[k-1]\\
    (k_1'''\dots k_u''')\subset[k-1]
    }}
    a_{k_1\dots k_s k}
    a_{k_1'\dots k_r' k} 
    a_{k_1''\dots k_t'' k}
    a_{k_1'''\dots k_u''' k} E(\tilde Z_{k_1\dots k_s}\tilde Z_{k_1'\dots k_r'}\tilde Z_{k_1''\dots k_t''}\tilde Z_{k_1'''\dots k_u'''}) 
     \right\}\\
     &\leq& 
     \frac{\nu_4^{4S}}{v_\Gamma^2} 
    \left(\sum_{s,r,t,u}^{{S}\wedge k} 
    \sum_{\substack{
    (k_1\dots k_s)\subset[k-1] \\ 
    (k_1'\dots k_r')\subset[k-1]\\
    (k_1''\dots k_t'')\subset[k-1]\\
    (k_1'''\dots k_u''')\subset[k-1]
    }}
    |a_{k_1\dots k_s k}|
    |a_{k_1'\dots k_r' k}| 
    |a_{k_1''\dots k_t'' k}|
    |a_{k_1'''\dots k_u''' k}|  
     \right)\\
     &\leq& 
     \frac{\nu_4^{4S} \bar{a}^4}{v_\Gamma^2} 
    \left(\sum_{\substack{W_1,W_2,W_3,W_4\subset[S]: \\
    W_1\cap W_2 \cap W_3 \cap W_4 \neq \varnothing
    }}
    \indicator\{|a_{W_1}|\neq 0\}
    \indicator\{|a_{W_2}|\neq 0\} 
    \indicator\{|a_{W_3}|\neq 0\}
    \indicator\{|a_{W_4}|\neq 0\}  
     \right)\\
     &\leq& 
     \frac{\nu_4^{4S} \bar{a}_m^4 (\B S)^4 m}{v_\Gamma^2} \ =\ o(1). 
\end{eqnarray*}

Combining results in steps I--III and Proposition~\ref{prop::martingale_clt}, we prove the results in Theorem~\ref{thm::Gamma-CLT}.
\qed

\section{Proofs}\label{sec::proofs}
\subsection{Two lemmas for proving the theorems}
\label{sec::lemma}
We first introduce two lemmas to simplify the proofs of the theorems.
Recall that in the bipartite graph with adjacency matrix $\bG$, we use $\cG_{+i}$ to denote the subset of indices of intervention units connected to outcome unit $i$, with cardinality $\G_{+i} = |\cG_{+i}|$. 
For ease of notation, in Sections~\ref{sec::proofs} and~\ref{sec::psd_lambda} of the supplementary material, we use $\Smaxsupp$ to denote $\maxi\G_{+i}$, the maximum degree of outcome units, and use $\Dmaxsupp$ to denote $\max_{1\leq k\leq m}\G_{k+}$, the maximum degree of intervention units. Symmetrically, we use $\cG_{k+}$ to denote the subset of indices of outcome units connected to intervention unit $k$ with cardinality $\G_{k+} = |\cG_{k+}|$. We use $T_i= \prod_{k\in \mathcal{G}_{+i}} Z_k$ and $C_i= \prod_{k\in \mathcal{G}_{+i}} (1-Z_k)$ to denote the indicator that all intervention units that outcome unit $i$ connects to were assigned to the treatment and the control, respectively, as introduced in the beginning of Section \ref{sec::point-estimator}.

\begin{lemma}
\label{lemma::double_sum_ab_rate}
For any two vectors $\{a_i\}_{i=1}^{n}$ and $\{b_i\}_{i=1}^{n}$, we have
\begin{eqnarray*}
    \left|n^{-2}\sumij a_ib_j \left(p^{-|\cGijcap|} - 1\right)\right| &\leq & n^{-1}\left(p^{-\Smaxsupp} - 1\right) \left(\maxi |a_i|\right)\left(\maxi |b_i|\right) \Smaxsupp \Dmaxsupp.
\end{eqnarray*}
\end{lemma}

\begin{proof}[Proof of Lemma~\ref{lemma::double_sum_ab_rate}]
$p^{-|\cGijcap|} - 1$ is nonzero if and only if $\cGijcap \neq \varnothing$. For each outcome unit $i$, the number of intervention units it connects to is no larger than $\Smaxsupp$, and there are at most $\Dmaxsupp$ outcome units connected to each intervention unit. Therefore, for each $1\leq i\leq n$, 
$$\left|\sumj b_j \left(p^{-|\cGijcap|} - 1\right)\right| \leq \left(p^{-\Smaxsupp} - 1\right)\left(\maxi |b_i|\right)\Smaxsupp \Dmaxsupp,$$ 
thus 
$$\left|\sumij a_ib_j \left(p^{-|\cGijcap|} - 1\right)\right| \leq n\left(p^{-\Smaxsupp} - 1\right) \left(\maxi |a_i|\right)\left(\maxi |b_i|\right) \Smaxsupp \Dmaxsupp.$$
This implies the desired result. 
\end{proof}

\begin{lemma}
\label{lemma::mean_var_ab_rate}
We have
\begin{eqnarray}
    && \left|E\left\{n^{-2}\sumij \frac{T_iT_j a_ib_j \left(p^{-|\cGijcap|} - 1\right)}{p^{|\cGijcup|}}\right\}\right| \notag \\
    &\leq & n^{-1}\left(p^{-\Smaxsupp} - 1\right) \left(\maxi |a_i|\right)\left(\maxi |b_i|\right) \Smaxsupp \Dmaxsupp, \label{eqn::mean_ab_rate} 
\end{eqnarray}
and 
\begin{eqnarray}
    && \var\left\{n^{-2}\sumij \frac{T_iT_j a_ib_j \left(p^{-|\cGijcap|} - 1\right)}{p^{|\cGijcup|}}\right\} \notag \\
    &\leq &n^{-3}p^{-4\Smaxsupp} \left(\maxi a_i^2\right)\left(\maxi b_i^2\right) \Smaxsupp^3 \Dmaxsupp^3\left(p^{-\Smaxsupp} - 1\right)^2. \label{eqn::var_ab_rate}
\end{eqnarray}
Similar results also hold for the quantities defined by $C_i$'s. 
\end{lemma}

\begin{proof}[Proof of Lemma~\ref{lemma::mean_var_ab_rate}]
By $E(T_iT_j)=p^{|\cGijcup|}$, we have
\begin{eqnarray*}
    E\left\{n^{-2}\sumij \frac{T_iT_j a_ib_j \left(p^{-|\cGijcap|} - 1\right)}{p^{|\cGijcup|}}\right\} &=& n^{-2}\sumij a_ib_j \left(p^{-|\cGijcap|}-1\right).
\end{eqnarray*}
Equation~\eqref{eqn::mean_ab_rate} holds by Lemma~\ref{lemma::double_sum_ab_rate}.

Next, we prove~\eqref{eqn::var_ab_rate}. Recall $(\Lambda_1)_{i,j}=p^{-|\cGijcap|} - 1$. We have
\begin{eqnarray}
    && \var\left\{n^{-2}\sumij \frac{T_iT_j a_i b_j \left(p^{-|\cGijcap|} - 1\right)}{p^{|\cGijcup|}}\right\} \notag \\
    & = &
    n^{-4}\sumij \sumu \sumv \frac{\cov(T_iT_j,T_uT_v) a_ib_ja_ub_v  (\Lambda_1)_{i,j} (\Lambda_1)_{u,v}}{p^{|\cGijcup|+|\cG_{+u} \cup \cG_{+v}|}} \notag \\
    & \leq &
    n^{-4}p^{-4\Smaxsupp} \sumij \sumu \sumv |\cov(T_iT_j,T_uT_v) a_ib_ja_ub_v  (\Lambda_1)_{i,j} (\Lambda_1)_{u,v}|. \label{eqn::var_ab_rate_proof_eqn1}
\end{eqnarray}
If $(\cGijcup) \cap (\cG_{+u} \cup \cG_{+v}) = \varnothing$, then $T_iT_j$ and $T_uT_v$ are independent, thus $\cov(T_iT_j,T_uT_v) =0$. Therefore, $\cov(T_iT_j,T_uT_v)(\Lambda_1)_{i,j} (\Lambda_1)_{u,v}$ is nonzero if and only if $\cGijcup \neq \varnothing$, $\cG_{+u} \cup \cG_{+v} \neq \varnothing$, and $(\cGijcup) \cap (\cG_{+u} \cup \cG_{+v}) \neq \varnothing$. Without loss of generality, assume that $(\cGijcup) \cap \cG_{+u} \neq \varnothing$, we have
\begin{eqnarray*}
    && \sumij \sumu \sumv |\cov(T_iT_j,T_uT_v) a_ib_ja_ub_v  (\Lambda_1)_{i,j} (\Lambda_1)_{u,v}| \\
    &\leq & \sumij |a_ib_j| (\Lambda_1)_{i,j} \sumu \sumv |\cov(T_iT_j,T_uT_v)a_ub_v(\Lambda_1)_{u,v}| \\
    &\leq & \sumij |a_ib_j| (\Lambda_1)_{i,j} \sumu |a_u|\sumv |\cov(T_iT_j,T_uT_v)b_v(\Lambda_1)_{u,v}| \\
    &\leq & \sumij |a_ib_j| (\Lambda_1)_{i,j} \sumu |a_u| \left(p^{-\Smaxsupp} - 1\right) \left(\max_{1\leq v\leq n} b_v\right) \Smaxsupp \Dmaxsupp \indicator \{(\cGijcup) \cap \cG_{+u} \neq \varnothing\}\\
    &\leq & \sumij |a_ib_j| (\Lambda_1)_{i,j} \left(p^{-\Smaxsupp} - 1\right) \left(\max_{1\leq u\leq n} a_u\right)\left(\max_{1\leq v\leq n} b_v\right) \Smaxsupp^2 \Dmaxsupp^2 \\
    &\leq & n\left(p^{-\Smaxsupp} - 1\right)^2 \left(\max_{1\leq u\leq n} a_u^2\right)\left(\max_{1\leq v\leq n} b_v^2\right) \Smaxsupp^3 \Dmaxsupp^3,
\end{eqnarray*}
where the third inequality follows from $(\cGijcup) \cap \cS_u \neq \varnothing$ and a similar argument in the proof of Lemma~\ref{lemma::double_sum_ab_rate} that for each $u$, $|\sumv b_v (\Lambda_1)_{u,v}| \leq (p^{-\Smaxsupp} - 1)(\max_{1\leq v\leq n} b_v)\Smaxsupp \Dmaxsupp$, the forth inequality follows from the fact that $u$ has to be connected to either $i$ of $j$, and the total number of $u$ such that $\indicator\{(\cGijcup) \cap \cG_{+u} \neq \varnothing\}$ is nonzero is no larger than $\Smaxsupp \Dmaxsupp$, and the last equality follows from Lemma~\ref{lemma::double_sum_ab_rate}. Plugging in back to~\eqref{eqn::var_ab_rate_proof_eqn1} gives the second inequality in Lemma~\ref{lemma::mean_var_ab_rate}. 
\end{proof}

\subsection{Proof of Theorem~\ref{thm::consistency}}
By the definitions of $\hat\mu_1$ and $\hat\mu_0$, we have:
\begin{eqnarray}
    \hat\mu_1 - \mu_1 &=& n^{-1} \sumi \frac{T_i (Y_i - \mu_1)}{p^{\G_{+i}}} \Big/ n^{-1}\sumi 
    \frac{T_i}{p^{\G_{+i}}}, \label{eqn::center-mu-1}\\ 
    \hat\mu_0 - \mu_0 &=& n^{-1}\sumi \frac{C_i (Y_i - \mu_0)}{(1-p)^{\G_{+i}}} \Big/ n^{-1}\sumi \frac{C_i}{(1-p)^{\G_{+i}}}.
    \label{eqn::center-mu-0}
\end{eqnarray}
We first show $\hat\mu_1$ is consistent to $\mu_1$ by showing the numerator of $\hat\mu_1-\mu_1$ in \eqref{eqn::center-mu-1} converges in probability to 0 and the denominator \eqref{eqn::center-mu-1} converges in probability to 1. The numerator of $\hat\mu_1-\mu_1$ in \eqref{eqn::center-mu-1} has mean zero and variance equal to
\allowdisplaybreaks
\begin{eqnarray*}
    && \var\left\{n^{-1} \sumi \frac{T_i (Y_i-\mu_1)}{p^{G_{+i}}}\right\} \\
    &=& E\left(\left[n^{-1}\sumi \frac{\left\{\Yt{i} - \mu_1\right\}\allt}{p^{G_{+i}}}\right]^2\right) \\
    &=& 
    n^{-2}\sumij\frac{1}{p^{G_{+i}+\G_{+j}}}E\left[\left\{\Yt{i} - \mu_1\right\}\left\{\Yt{j} - \mu_1\right\} \allt\alltj \right]\\
    &=&
    n^{-2}\sumij\frac{\left\{\Yt{i} - \mu_1\right\}\left\{\Yt{j} - \mu_1\right\} p^{|\cGijcup|}}{p^{G_{+i}+\G_{+j}}} \\
    &=&
    n^{-2}\sumij \left\{\Yt{i} - \mu_1\right\}\left\{\Yt{j} - \mu_1\right\} p^{-|\cGijcap|} \\
    &=&
    n^{-2}\sumij \left\{\Yt{i} - \mu_1\right\}\left\{\Yt{j} - \mu_1\right\} \left(p^{-|\cGijcap|}-1\right) \\
    &\leq & n^{-1}\left(p^{-\Smaxsupp} - 1\right) \left(\maxi a_i\right)\left(\maxi b_i\right) \Smaxsupp \Dmaxsupp,
\end{eqnarray*}
where the second-to-last equality follows from the fact that $\sumij\{\Yt{i} - \mu_1\}\{\Yt{j} - \mu_1\}=0$, and the last inequality follows from Lemma~\ref{lemma::double_sum_ab_rate}. Therefore, the numerator of $\hat\mu_1-\mu_1$ in \eqref{eqn::center-mu-1} converges in probability to 0 by Chebyshev's inequality. Similarly, the denominator of $\hat\mu_1-\mu_1$ in \eqref{eqn::center-mu-1} has mean 1 and variance converging in probability to 0. This concludes the proof of $\hat\mu_1$ converges in probability to $\mu_1$. Analogously, $\hat\mu_0$ converges in probability to $\mu_0$, which concludes the proof of Theorem~\ref{thm::consistency}.
\qed

\subsection{Proof of Theorem~\ref{thm::asym_dist}}

We divide the proof of Theorem \ref{thm::asym_dist} into three steps. In step I, we derive the asymptotic variance of the proposed estimator $\hat\mu_1$. In step II, we give an alternative representation of the numerator of $\hat\mu_1 - \mu_1$ in \eqref{eqn::center-mu-1} in the form of a random polynomial, as introduced in \eqref{eqn::Gamma}. In step III, we apply Theorem \ref{thm::Gamma-CLT} to this random polynomial representation to establish the asymptotic normality of the estimator. We can follow similar steps to analyze a Horvitz--Thompson estimator.    

\noindent \textbf{Step I.}
We first compute the asymptotic variance of the proposed estimators $\hat\mu_1$. The denominator of $\hat\mu_1 - \mu_1$ in \eqref{eqn::center-mu-1} converges in probability to 1. By Slutsky's theorem, we have
\begin{eqnarray*}
    && \avar(\hat\mu_1) \\
    &=& \var\left[n^{-1} \sumi \frac{\left\{\Yt{i} - \mu_1\right\}\allt}{p^{G_{+i}}}\right] \\
    &=& E\left(\left[n^{-1}\sumi \frac{\left\{\Yt{i} - \mu_1\right\}\allt}{p^{G_{+i}}}\right]^2\right) \\
    &=& 
    n^{-2}\sumij\frac{1}{p^{G_{+i}+\G_{+j}}}E\left[\left\{\Yt{i} - \mu_1\right\}\left\{\Yt{j} - \mu_1\right\} \allt\alltj \right]\\
    &=&
    n^{-2}\sumij\frac{\left\{\Yt{i} - \mu_1\right\}\left\{\Yt{j} - \mu_1\right\} p^{|\cGijcup|}}{p^{G_{+i}+\G_{+j}}} \\
    &=&
    n^{-2}\sumij {\left\{\Yt{i} - \mu_1\right\}\left\{\Yt{j} - \mu_1\right\} p^{-|\cGijcap|}} \\
    &=& n^{-2}\sumij \left\{\Yt{i} - \mu_1\right\}\left\{\Yt{j} - \mu_1\right\} (\Lambda_1)_{i,j}.
\end{eqnarray*}
By symmetry, we have
\begin{eqnarray*}
    \avar(\hat\mu_0) &=& n^{-2}\sumij\left\{\Yc{i} - \mu_0\right\}\left\{\Yc{j} - \mu_0\right\} (\Lambda_0)_{i,j}. 
\end{eqnarray*}
Next, we compute the asymptotic covariance between $\hat\mu_1$ and $\hat\mu_0$:
\begin{eqnarray*}
    \acov(\hat\mu_1, \hat\mu_0) 
    &=& n^{-2} E\left[
    \sumi \frac{\left\{\Yt{i} - \mu_1\right\}\allt}{p^{G_{+i}}}, 
    \sumi \frac{\left\{\Yc{i} - \mu_0\right\}\allc}{(1-p)^{G_{+i}}}\right] \\
    &=& n^{-2} \sumij E\left[ \frac{\left\{\Yt{i} - \mu_1\right\}\left\{\Yc{j} - \mu_0\right\}\allt\allcj}{p^{G_{+i}}(1-p)^{\G_{+j}}} \right] \\
    &=& n^{-2} \sumij \left\{\Yt{i} - \mu_1\right\}\left\{\Yc{j} - \mu_0\right\}\indicator\{\cGijcap = \varnothing\}\\
    &=& -n^{-2}\sumij \left\{\Yt{i} - \mu_1\right\}\left\{\Yc{j} - \mu_0\right\}\indicator\{\cGijcap \neq  \varnothing\}. 
\end{eqnarray*}
Combining the results, we have
\begin{eqnarray*}
    \avar(\hat\tau) &=& \avar(\hat\mu_1) + \avar(\hat\mu_2) - 2\acov(\hat\mu_1, \hat\mu_2)\\
    &=& n^{-2}\sumij \left\{\Yt{i} - \mu_1\right\}\left\{\Yt{j} - \mu_1\right\} (\Lambda_1)_{i,j} + n^{-2}\sumij \left\{\Yc{i} - \mu_0\right\}\left\{\Yc{j} - \mu_0\right\} (\Lambda_0)_{i,j}\\
    &&+ 2n^{-2}\sumij \left\{\Yt{i} - \mu_1\right\}\left\{\Yc{j} - \mu_0\right\} (\Lambda_\tau)_{i,j} \\
    &=& n^{-2}\left\{\tilde\bY(\bOne)^{\T} \Lambda_1 \tilde\bY(\bOne) + \tilde\bY(\bZero)^{\T} \Lambda_0 \tilde\bY(\bZero) +2 \tilde\bY(\bOne)^{\T} \Lambda_{\tau} \tilde\bY(\bZero) \right\}.
\end{eqnarray*}    

We then apply Theorem~\ref{thm::Gamma-CLT} following two further steps.

\noindent \textbf{Step II.}
We first give an alternative representation of the numerator of $\hat\mu_1 - \mu_1$ in \eqref{eqn::center-mu-1}. Introduce the notation $\tilde{Z}_k = Z_k - p$ to write
\begin{eqnarray}
    &&
    n^{-1}\sumi\frac{\allt \{\Yt{i} - \mu_1\}}{p^{G_{+i}}}\notag\\
    &=& 
    \sum_{k_1=1}^{m} \frac{Z_{k_1}}{n p} \sumi \indicator\{\G_{+i} = 1\} \G_{k_1i}\{\Yt{i} - \mu_1\}\notag\\
    && + 
    \sum_{(k_1 k_2)\subset[m]}\frac{Z_{k_1}Z_{k_2}}{n p^2} \sumi \indicator \{\G_{+i} = 2\} \G_{k_1 i}\G_{k_2 i}\{\Yt{i} - \mu_1\}\notag\\
    && + \cdots \notag\\
    && + 
    \sum_{(k_1 \dots k_\Smaxsupp)\subset[m]}\frac{Z_{k_1}\cdots Z_{k_\Smaxsupp}}{n p^\Smaxsupp} \sumi \indicator\{G_{+i} = \Smaxsupp\}G_{k_1 i}\cdots G_{k_\Smaxsupp i}\{\Yt{i} - \mu_1\}\notag\\
    &=&
    \sum_{k_1=1}^{m}\frac{\tilde{Z}_{k_1} + p}{n p} \sumi \indicator\{G_{+i} = 1\} \G_{k_1i} \{\Yt{i} - \mu_1\}\notag\\
    && + 
    \sum_{(k_1 k_2)\subset[m]}\frac{(\tilde Z_{k_1} + p)(\tilde Z_{k_2} + p)}{n p^2} \sumi \indicator\{G_{+i} = 2\} \G_{k_1i} \G_{k_2i} \{\Yt{i} - \mu_1\}\notag\\
    && + \cdots \notag\\
    && + 
    \sum_{(k_1 \dots k_\Smaxsupp)\subset[m]}\frac{\Pi_{s=1}^{\Smaxsupp}(\tilde{Z}_{k_{s}} + p)}{n p^\Smaxsupp} \sumi \indicator\{G_{+i} = \Smaxsupp\} \G_{k_1i} \cdots \G_{k_\Smaxsupp i} \{\Yt{i} - \mu_{1}\}. \label{eqn::sum-mu1}
\end{eqnarray}
By binomial expansion, for any $s\in[\Smaxsupp]$, we have 
\begin{eqnarray}
    &&\sum_{(k_1 \dots k_s)\subset[m]}\frac{(\tilde{Z}_{k_1} + p)\cdots (\tilde{Z}_{k_{s}}+p)}{n p^{s}} \sumi \indicator\{G_{+i} = s\} \G_{k_1i} \cdots \G_{k_{s}i} \{\Yt{i} - \mu_1\} \notag \\
    &=& \frac{1}{s!} \sum_{k_1 \neq \dots \neq k_s\in[m]}\frac{(\tilde{Z}_{k_1} + p)\cdots (\tilde{Z}_{k_{s}}+p)}{n p^{s}} \sumi \indicator\{G_{+i} = s\} \G_{k_1i} \cdots \G_{k_{s}i} \{\Yt{i} - \mu_1\} \notag \\
    &=& \frac{1}{s!} {s\choose 1}\sum_{k_1\in[m]} \frac{\tilde{Z}_{k_1}}{n p} \sumi \indicator\{G_{+i} = s\} \G_{k_1i} \{\Yt{i} - \mu_1\} \sum_{\substack{k_{2}\neq\cdots\neq k_{s}\in[m], \\ k_{u}\neq k_1, \forall 1<u\leq s}} \G_{k_2i} \cdots \G_{k_{s}i} \notag \\
    &&+ \cdots \notag \\
    &&+ \frac{1}{s!} {s\choose \ell}\sum_{k_1 \neq \dots \neq k_\ell} \frac{\tilde{Z}_{k_1} \cdots \tilde{Z}_{k_\ell}}{n p^\ell} \sumi \indicator\{G_{+i} = s\} \G_{k_1i} \cdots \G_{k_{\ell}i} \{\Yt{i} - \mu_1\} \sum_{\substack{k_{\ell+1}\neq\cdots\neq k_{s}\in[m], \\ k_{u}, \ldots, k_\ell, \forall \ell<u\leq s}} \G_{k_{\ell+1}i} \cdots \G_{k_{s}i} \notag \\
    &&+ \cdots \notag \\ 
    &&+ \frac{1}{s!} {s \choose s-1}\sum_{k_1 \neq \dots \neq k_{s-1} \in[m]} \frac{\tilde{Z}_{k_1} \cdots \tilde{Z}_{k_{s-1}}}{n p^{s-1}} \sumi \indicator\{G_{+i} = s\} \G_{k_1i} \cdots \G_{k_{s-1}i} \{\Yt{i} - \mu_1\} \sum_{\substack{k_s\in[m], \\k_s\neq k_1, \ldots, k_{s-1}}} \G_{k_si} \notag \\ 
    &&+ \frac{1}{s!} \sum_{k_1 \neq \dots \neq k_s\in[m]} \frac{\tilde{Z}_{k_1} \cdots \tilde{Z}_{k_s}}{n p^{s}} \sumi \indicator\{G_{+i} = s\} \G_{k_1i} \cdots \G_{k_{s}i} \{\Yt{i} - \mu_1\} \label{eqn::app_clt_proof_eqn1}.
\end{eqnarray}
For the $\ell$-th term, we have
\begin{eqnarray*}
    && \frac{1}{s!} {s\choose \ell}\sum_{k_1 \neq \dots \neq k_\ell\in[m]} \frac{\tilde{Z}_{k_1} \cdots \tilde{Z}_{k_\ell}}{n p^\ell} \sumi \indicator\{G_{+i} = s\} \G_{k_1i} \cdots \G_{k_{\ell}i} \{\Yt{i} - \mu_1\} \sum_{\substack{k_{\ell+1}\neq\cdots\neq k_{s}\in[m], \\ k_{u}\neq k_1, \ldots, k_\ell, \forall \ell<u\leq s}} \G_{k_{\ell+1}i} \cdots \G_{k_{s}i} \\
    &=& \frac{1}{s!} {s\choose \ell}\sum_{k_1 \neq \dots \neq k_\ell\in[m]} \frac{\tilde{Z}_{k_1} \cdots \tilde{Z}_{k_\ell}}{n p^\ell} \sumi \indicator\{G_{+i} = s\} \G_{k_1i} \cdots \G_{k_{\ell}i} \{\Yt{i} - \mu_1\} (s-\ell)! \\
    &=& \frac{1}{s!} {s\choose \ell}(s-\ell)! \ell! \sum_{(k_1  \dots  k_\ell)\subset[m]} \frac{\tilde{Z}_{k_1} \cdots \tilde{Z}_{k_\ell}}{n p^\ell} \sumi \indicator\{G_{+i} = s\} \G_{k_1i} \cdots \G_{k_{\ell}i} \{\Yt{i} - \mu_1\} \\
    &=& \sum_{(k_1  \dots  k_\ell)\subset[m]} \frac{\tilde{Z}_{k_1} \cdots \tilde{Z}_{k_\ell}}{n p^\ell} \sumi \indicator\{G_{+i} = s\} \G_{k_1i} \cdots \G_{k_{\ell}i} \{\Yt{i} - \mu_1\}.
\end{eqnarray*}
Plugging back to~\eqref{eqn::app_clt_proof_eqn1}, we have
\begin{eqnarray*}
    && \sum_{(k_1 \dots k_s)\subset[m]}\frac{(\tilde{Z}_{k_1} + p)\cdots (\tilde{Z}_{k_{s}}+p)}{n p^{s}} \sumi \indicator\{G_{+i} = s\} \G_{k_1i} \cdots \G_{k_{s}i} \{\Yt{i} - \mu_1\} \\
    &=& \sum_{\ell=1}^{s} \sum_{(k_1 \dots k_\ell)\subset [m]} \frac{\tilde{Z}_{k_1} \cdots \tilde{Z}_{k_\ell}}{n p^\ell} \sumi \indicator\{G_{+i} = s\} \G_{k_1i} \cdots \G_{k_{\ell}i} \{\Yt{i} - \mu_1\},
\end{eqnarray*}
thus, the summation in~\eqref{eqn::sum-mu1} is equal to
\begin{eqnarray*}
    &&\sum_{k_1\in[m]} \frac{\tilde{Z}_{k_1}}{n p} \sumi \G_{k_1i} \{\Yt{i} - \mu_1\} \sum_{s=1}^\Smaxsupp \indicator\{G_{+i} = s\}  \\
    && + \sum_{(k_1 k_2)\subset[m]} \frac{\tilde{Z}_{k_1}\tilde{Z}_{k_2}}{n p^2} \sumi \G_{k_1i} \G_{k_2i} \{\Yt{i} - \mu_1\} \sum_{s=2}^\Smaxsupp \indicator\{G_{+i} = s\} \\
    && + \cdots \\
    && + \sum_{(k_1 \dots k_\Smaxsupp)\subset[m]} \frac{\tilde{Z}_{k_1}  \cdots  \tilde{Z}_{k_\Smaxsupp}}{n p^\Smaxsupp} \sumi \G_{k_1i} \cdots \G_{k_\Smaxsupp i} \{\Yt{i} - \mu_1\}  \sum_{s=\Smaxsupp}^\Smaxsupp \indicator\{G_{+i} = s\}. 
\end{eqnarray*}
By symmetry, the numerator of $\hat\mu_0 - \mu_0$ in \eqref{eqn::center-mu-0} equals
\begin{eqnarray*}
    && n^{-1}\sumi\frac{\allt\{\Yc{i} - \mu_0\}}{n (1-p)^{G_{+i}}} \\
    &=& \sum_{k_1\in[m]} -\frac{\tilde{Z}_{k_1}}{n (1-p)} \sumi \G_{k_1i} \{\Yc{i} - \mu_0\} \sum_{s=1}^\Smaxsupp \indicator\{G_{+i} = s\}  \\
    && + \sum_{(k_1 k_2)\subset[m]} (-1)^{2} \frac{\tilde{Z}_{k_1}\tilde{Z}_{k_2}}{n (1-p)^2} \sumi \G_{k_1i} \G_{k_2i} \{\Yc{i} - \mu_0\} \sum_{s=2}^\Smaxsupp \indicator\{G_{+i} = s\} \\
    && + \cdots \\
    && + \sum_{(k_1 \dots k_\Smaxsupp)\subset[m]} (-1)^{\Smaxsupp} \frac{\tilde{Z}_{k_1}  \cdots  \tilde{Z}_{k_\Smaxsupp}}{n (1-p)^\Smaxsupp} \sumi \G_{k_1i} \cdots \G_{k_\Smaxsupp i} \{\Yc{i} - \mu_0\}  \sum_{s=\Smaxsupp}^\Smaxsupp \indicator\{G_{+i} = s\}. \\
\end{eqnarray*}
Define
\begin{eqnarray*}
    a_{1,k_1\cdots k_{\ell}} &=& \frac{1}{n p^{\ell}}\sumi \G_{k_1i} \cdots \G_{k_{\ell}i} \{\Yt{i} - \mu_1\} \sum_{s=\ell}^\Smaxsupp \indicator\{G_{+i} = s\}, \\
    a_{0,k_1\cdots k_{\ell}} &=& \frac{(-1)^{\ell}}{n (1-p)^{\ell}}\sumi \G_{k_1i} \cdots \G_{k_{\ell}i} \{\Yc{i} - \mu_0\} \sum_{s=\ell}^\Smaxsupp \indicator\{G_{+i} = s\}.
\end{eqnarray*}
To summarize, we have shown that the numerator of $\hat\mu_z - \mu_z$ in \eqref{eqn::center-mu-0} and $\eqref{eqn::center-mu-1}$ is equal to
$$\sum_{\ell=1}^{\Smaxsupp}\sum_{(k_1 \cdots k_{\ell})\subset[m]} a_{z,k_1\cdots k_{\ell}} \tilde Z_{k_1}\cdots \tilde Z_{k_{\ell}}$$
for $z=1,0$.

\noindent \textbf{Step III.}
We now consider any linear combination of the numerators of $\hat\mu_1-\mu_1$ in \eqref{eqn::center-mu-1} and $\hat\mu_0-\mu_0$ in \eqref{eqn::center-mu-0}. We show it can be reformulated in the form of $\Gamma$ defined in~\eqref{eqn::Gamma}. Consider any $c_1$ and $c_0$ that has $c_1^2 + c_0^2 = 1$. Define
\begin{eqnarray*}
    a_{k_1\cdots k_{\ell}} 
    &=& 
    c_1a_{1, k_1\cdots k_{\ell}} 
    + 
    c_0a_{0, k_1\cdots k_{\ell}}.
\end{eqnarray*}
Then we can write
\begin{eqnarray}
    n^{-1}\sumi \left[\frac{c_1T_i\{\Yt{i}-\mu_1\}}{p^{G_{+i}}} + \frac{c_0C_i\{\Yc{i}-\mu_0\}}{(1-p)^{G_{+i}}}\right] 
    &=&
    \sum_{\ell=1}^{\Smaxsupp}\sum_{(k_1 \cdots k_{\ell})\subset[m]} a_{k_1\cdots k_{\ell}} \tilde Z_{k_1}\cdots \tilde Z_{k_{\ell}}.\label{eqn::est-c1c0}
\end{eqnarray}
We will apply Theorem~\ref{thm::Gamma-CLT} to establish a central limit theorem for~\eqref{eqn::est-c1c0}. We check the two conditions required in Theorem~\ref{thm::Gamma-CLT}.

We first show the boundedness of $a$'s. Note that
\begin{eqnarray*}
    a_{k_1\cdots k_{\ell}} &=& 
    \sumi \G_{k_1i} \cdots \G_{k_{\ell}i} \left[
    \frac{c_1\{\Yt{i} - \mu_1\}}{n p^{\ell}} 
    +
    \frac{(-1)^\ell c_0\{\Yc{i} - \mu_0\}}{n (1-p)^{\ell}} \right] \sum_{s=\ell}^\Smaxsupp \indicator\{G_{+i} = s\}.
\end{eqnarray*}

The summand indexed by $i$ is nonzero only if outcome unit $i$ connects with intervention unit $k_1,\ldots,k_\ell$. By Assumption~\ref{assump::Dmax}, for each $k_1,\ldots,k_\ell$, we have at most $\Dmaxsupp$ such outcome units. Hence we obtain 
\begin{eqnarray*}
    |a_{k_1\cdots k_{\ell}}|
    &\leq &
    \frac{\Dmaxsupp\maxi\{|\Yt{i} - \mu_1|, |\Yc{i} - \mu_0|\}}{n}\left\{p^{-\Smaxsupp} + (1-p)^{-\Smaxsupp}\right\}:=\bar{a}_m. 
\end{eqnarray*}
    
Second, we verify the limited overlapping condition $\sum_{(k_1 \cdots k_s)\subset[m]\backslash \{k\}}\indicator\{|a_{kk_1\ldots k_s}|\neq 0 \} \leq \B$. For any $(k_1\cdots k_s)\subset[m]$ and $k$, we have $\indicator\{|a_{kk_1\ldots k_s}|\neq 0 \}=\indicator\{\exists\ i, \text{ such that }\G_{k_1i} \cdots \G_{k_si} \G_{ki} = 1\}$, which is nonzero if and only if $k_1,\ldots, k_s$ are all connected to intervention unit $k$. Therefore, 
\begin{eqnarray*}
    && \sum_{(k_1\cdots k_s)\subset[m]\backslash \{k\}}\indicator\{|a_{kk_1\ldots k_s}|\neq 0 \} \\
    &\leq& 
    \sum_{(k_1 \cdots k_s)\subset[m]\backslash \{k\}}\indicator\{k_1,\ldots, k_s\textup{ are all connected to intervention unit }k\} \ \leq\ {\B}^{s},
\end{eqnarray*}
where the last inequality holds because by Assumption~\ref{assump::sparse}, there are at most $\B$ intervention units connected to intervention unit $k$, thus the number of combinations $(k_1,\ldots,k_s)$ such that all of them are connected to $k$ is upper bounded by ${\B\choose{s}} \leq \B^{s}$.

Therefore, by step II, we conclude that the numerators of $\hat\mu_1 - \mu_1$ and $\hat\mu_0 - \mu_0$ in \eqref{eqn::center-mu-1} and \eqref{eqn::center-mu-0} converge jointly to a bivariate standard normal distribution, after standardization via
\begin{eqnarray*}
    \begin{pmatrix}
        n^{-2}\tilde\bY(\bOne)^{\T} \Lambda_1\tilde\bY(\bOne) 
        & 
        -n^{-2}\tilde\bY(\bOne)^{\T} \Lambda_{\tau} \tilde\bY(\bZero)\\
        -n^{-2}\tilde\bY(\bZero)^{\T} \Lambda_{\tau} \tilde\bY(\bOne)
        &
        n^{-2}\tilde\bY(\bZero)^{\T} \Lambda_0
        \tilde\bY(\bZero)
    \end{pmatrix}. 
\end{eqnarray*}

Moreover, the denominators of $\hat\mu_1 - \mu_1$ and $\hat\mu_0 - \mu_0$ in \eqref{eqn::center-mu-1} and \eqref{eqn::center-mu-0} are converging in probability to 1, thus the asymptotic distribution in Theorem~\ref{thm::asym_dist} holds by Slutsky's Theorem.
\qed

\subsection{Proof of the result in Remark~\ref{rmk::regularity_condition_super_population}}

We show that the following results hold almost surely:
\begin{eqnarray}
    n^{-1}\tilde\bY(\bOne)^{\T} \Lambda_1\tilde\bY(\bOne)
    -
    n^{-1}\sigma_1^2\sum_{i=1}^n (p^{-\G_{+i}} - 1)
     \rightarrow 
    0, \label{eqn::v1-special}\\
    n^{-1}\tilde\bY(\bZero)^{\T} \Lambda_0\tilde\bY(\bZero)
    - 
    n^{-1}\sigma_0^2\sum_{i=1}^n \{(1-p)^{-\G_{+i}} - 1\}
     \rightarrow 
    0, \label{eqn::v0-special}\\
    n^{-1}\tilde\bY(\bOne)^{\T} \Lambda_{\tau} \tilde\bY(\bZero)
     \rightarrow 
    0. \label{eqn::v10-special}
\end{eqnarray}

Let $\Pi_n = I_n - n^{-1}1_n1_n^{\T}$. For \eqref{eqn::v1-special}, we have  
\begin{eqnarray*}
    \frac{1}{n}\tilde\bY(\bOne)^{\T} \Lambda_1\tilde\bY(\bOne)
    & = &
    \frac{1}{n}\bY(\bOne)^{\T} \Pi_n \Lambda_1 \Pi_n \bY(\bOne),
\end{eqnarray*}
which, in distribution, is equivalent to $\sigma^2n^{-1}\sum_{i=1}^n \varrho_i \xi_i^2$, where $\varrho_i$ is the $i$-th eigenvalue of $\Pi_n \Lambda_1 \Pi_n$ and $\xi_i$ are i.i.d. standard Gaussian variables. By the strong law of large numbers, 
\begin{eqnarray*}
    \frac{\sigma^2}{n}\sum_{i=1}^n \varrho_i (\xi_i^2 - 1)
    \rightarrow 0, \quad \text{almost surely}.
\end{eqnarray*}
Now we have
\begin{eqnarray*}
    \frac{1}{n}\sum_{i=1}^n\varrho_i = \frac{1}{n}\operatorname{tr}(\Pi_n \Lambda_1 \Pi_n) = \frac{1}{n}\operatorname{tr}(\Lambda_1 \Pi_n)
    =
    \frac{1}{n}\operatorname{tr}(\Lambda_1) - \frac{1}{n^2} 1_n^{\T}\Lambda_1 1_n = \frac{1}{n}\sum_{i=1}^n (p^{-\G_{+i}} - 1) + o(1),
\end{eqnarray*}
where the last equality follows from the fact that the maximal degree $\Dmaxsupp$ is bounded. Therefore, we conclude~\eqref{eqn::v1-special}. The proof of~\eqref{eqn::v0-special} is similar and thus omitted. 

For~\eqref{eqn::v10-special}, by eigenvalue decomposition, we obtain that $n^{-1}\tilde\bY(\bOne)^{\T} \Lambda_{\tau} \tilde\bY(\bZero)$ is equivalent to $n^{-1}\sigma_1\sigma_0 \sum_{i=1}^n \nu_i \xi_i\xi_i'$ in distribution, where $\nu_i$'s are the eigenvalues of $\Pi_n\Lambda_\tau \Pi_n$, and $(\xi_i, \xi_i')_{i=1}^n$ are i.i.d. standard bivariate Gaussian vectors. By the strong law of large numbers, we have
\begin{align*}
    n^{-1}\tilde\bY(\bOne)^{\T} \Lambda_{\tau} \tilde\bY(\bZero) \rightarrow 0, \quad \text{almost surely}.
\end{align*}

The almost surely convergence in~\eqref{eqn::v1-special}--~\eqref{eqn::v10-special} ensures that 
\begin{eqnarray*}
    n v_n - \left[n^{-1}\sigma_1^2\sum_{i=1}^n (p^{-\G_{+i}} - 1) + n^{-1}\sigma_0^2\sum_{i=1}^n \{(1-p)^{-\G_{+i}} - 1\}\right] \rightarrow 0, \quad \text{almost surely}.
\end{eqnarray*} 
\qed

\subsection{Proof of Theorem~\ref{thm::var_est}}

\noindent \textbf{Part (a).} We first prove that $v_{n,\textup{UB}}\geq v_{n}$. Recall that $v_{n,\textup{UB}}=\{\plim(\hat v_1)^{1/2}+\plim(\hat v_0)^{1/2}\}^2$, by Cauchy--Schwarz  inequality, we have
\begin{eqnarray*}
    \avar(\hat\tau) &=& \avar(\hat\mu_1)+\avar(\hat\mu_0)-2\acov(\hat\mu_1,\hat\mu_0) \\
    &\leq& \plim(\hat v_1)+\plim(\hat v_0) + 2\plim(\hat v_1)^{1/2}\plim(\hat v_0)^{1/2} \ =\ \plim(\hat v)=v_{n}.
\end{eqnarray*}

\noindent \textbf{Part (b).} Next, we prove the convergence of $\hat v/\plim(\hat v)$. Recall that 
\begin{eqnarray*}
    \hat{v}_1 &=& n^{-2}\sumij \frac{T_iT_j(Y_i - \hat\mu_1)(Y_j - \hat\mu_1)(\Lambda_1)_{i,j}}{p^{|\cGijcup|}}, \\
    \hat{v}_0 &=& n^{-2}\sumij \frac{C_iC_j(Y_i - \hat\mu_0)(Y_j - \hat\mu_0)(\Lambda_0)_{i,j}}{(1-p)^{|\cGijcup|}},
\end{eqnarray*}
then we have $\hat v = (\hat{v}_1^{1/2} + \hat{v}_0^{1/2})^2$. We prove the convergence of $\hat v/\plim(\hat v)$ by showing that $\hat v_1/ \plim(\hat v_1) = 1+o_{P}(1)$ and $\hat v_0/ \plim(\hat v_0) = 1+o_{P}(1)$, where 
\begin{eqnarray*}
    \plim(\hat v_1) &=& \avar(\hat\mu_1) \ =\ n^{-2}\tilde\bY(\bOne)^{\T} \Lambda_1 \tilde\bY(\bOne), \\
    \plim(\hat v_0) &=& \avar(\hat\mu_0) \ =\ n^{-2}\tilde\bY(\bZero)^{\T} \Lambda_0 \tilde\bY(\bZero).
\end{eqnarray*}
Rewrite
\begin{eqnarray}
    \hat{v}_1 &=& n^{-2}\sumij \frac{T_iT_j\{Y_i - \mu_1 + (\mu_1 - \hat\mu_1)\}\{Y_j - \mu_1 + (\mu_1 - \hat\mu_1)\} (p^{-|\cGijcap|} - 1)}{p^{|S_i\cup S_j|}} \notag \\
    &=& n^{-2}\sumij \frac{T_iT_j(Y_i - \mu_1)(Y_j - \mu_1) (p^{-|\cGijcap|} - 1)}{p^{|\cGijcup|}} \label{eqn::var_est_proof_T1} \\
    &&+ 2(\mu_1 - \hat\mu_1) n^{-2}\sumij \frac{T_iT_j(Y_i - \mu_1) (p^{-|\cGijcap|} - 1)}{p^{|\cGijcup|}} \label{eqn::var_est_proof_T2} \\
    &&+ (\mu_1 - \hat\mu_1)^2 n^{-2}\sumij \frac{T_iT_j  (p^{-|\cGijcap|} - 1)}{p^{|\cGijcup|}}, \label{eqn::var_est_proof_T3}
\end{eqnarray}
and use $\cT_1, \cT_2, \cT_3$ to denote the three terms in~\eqref{eqn::var_est_proof_T1}--\eqref{eqn::var_est_proof_T3}, respectively. By the fact that $E(T_i T_j)=p^{|\cGijcup|}$, we have $E(\cT_1)=\plim(\hat v_1)$. The variance of $\cT_1$,
\begin{eqnarray*}
    \var(\cT_1) &\leq& n^{-3}p^{-4\Smaxsupp} [\maxi \{\Yt{i}-\mu_1\}^4] \Smaxsupp^3 \Dmaxsupp^3(p^{-\Smaxsupp} - 1)^2 \ =\  O_{P}(n^{-3}\Dmaxsupp^3)
\end{eqnarray*}
by Lemma~\ref{lemma::mean_var_ab_rate} when taking $a_i=b_i=\Yt{i}-\mu_1$. Thus,
\begin{eqnarray*}
    \cT_1 &=& E(\cT_1) + O_{P}\{\var(\cT_1)^{1/2}\} \ =\ \plim(\hat v_1) + O_{P}(n^{-3/2}\Dmaxsupp^{3/2}) \ =\ \plim(\hat v_1) + o_{P}(n^{-1}\Dmaxsupp).
\end{eqnarray*}
Similarly, by Lemma~\ref{lemma::mean_var_ab_rate}, we have
\begin{eqnarray*}
    E\left\{n^{-2}\sumij \frac{T_iT_j(Y_i - \mu_1) (p^{-|\cGijcap|} - 1)}{p^{|\cGijcup|}}\right\} &=& O_{P}(n^{-1}\Dmaxsupp), \\
    \var\left\{n^{-2}\sumij \frac{T_iT_j(Y_i - \mu_1) (p^{-|\cGijcap|} - 1)}{p^{|\cGijcup|}}\right\} &=& O_{P}(n^{-3}\Dmaxsupp^3),
\end{eqnarray*}
by taking $a_i=\Yt{i}-\mu_1$ and $b_i=1$. By the proof of Theorem~\ref{thm::asym_dist}, we have $\hat\mu_1 - \mu_1 = O_{P}(n^{-1/2}\Dmaxsupp^{1/2})$, and $\cT_2 = E(\cT_2)+O_{P}\{\var(\cT_2)^{1/2}\}$ gives us
\begin{eqnarray*}
    \cT_2 &=& O_{P}(n^{-1/2}\Dmaxsupp^{1/2})\cdot O_{P}(n^{-1}\Dmaxsupp) + O_{P}(\{n^{-1}\Dmaxsupp\cdot n^{-3}\Dmaxsupp^3\}^{1/2}) \ =\ o_{P}(n^{-1}\Dmaxsupp).
\end{eqnarray*}
Also, we have
\begin{eqnarray*}
    E\left\{n^{-2}\sumij \frac{T_iT_j  (p^{-|\cGijcap|} - 1)}{p^{|\cGijcup|}}\right\} &=& O_{P}(n^{-1}\Dmaxsupp), \\
    \var\left\{n^{-2}\sumij \frac{T_iT_j  (p^{-|\cGijcap|} - 1)}{p^{|\cGijcup|}}\right\} &=& O_{P}(n^{-3}\Dmaxsupp^3)
\end{eqnarray*}
by taking $(a_i,b_i)=(1,1)$ in Lemma~\ref{lemma::mean_var_ab_rate}. Again, we have
\begin{eqnarray*}
    \cT_3 &=& O_{P}(n^{-1}\Dmaxsupp)\cdot O_{P}(n^{-1}\Dmaxsupp) + O_{P}(\{n^{-2}\Dmaxsupp^2\cdot n^{-3}\Dmaxsupp^3\}^{1/2}) \ =\ o_{P}(n^{-1}\Dmaxsupp).
\end{eqnarray*}
Combining the three terms $\cT_1$--$\cT_3$, we have
\begin{eqnarray*}
    \hat v_1 &=& \plim(\hat v_1) + o_{P}(n^{-1}\Dmaxsupp).
\end{eqnarray*}
Under the regularity condition that the weighted covariance matrix of the potential outcomes $\Yt{i}$ and $\Yc{i}$ are non-degenerated, $\plim(\hat v_1)=O_{P}(n^{-1}\Dmaxsupp)$, thus $\hat v_1 / \plim(\hat v_1) = 1 + o_{P}(1)$. Analogously, $\hat v_0 / \plim(\hat v_0) = 1 + o_{P}(1)$. By the continuous mapping theorem, $\hat v / \plim(\hat v)$ converges in probability to 1.
\qed

\subsection{Proof of Lemma~\ref{lemma::L_beta1_beta0}}
Recall that we have
\begin{eqnarray*}
    n^2 v_n(\beta_1,\beta_0) &=& \{\tilde\bY(\bOne) - \bX \beta_1\}^{\T} \Lambda_1\{\tilde\bY(\bOne) - \bX \beta_1 \} - \{\tilde\bY(\bZero) - \bX \beta_0\}^{\T} \Lambda_0\{\tilde\bY(\bZero) - \bX \beta_0 \} \\
    &&+ 2 \{\tilde\bY(\bOne) - \bX \beta_1\}^{\T} \Lambda_{\tau} 
    \{\tilde\bY(\bZero) - \bX \beta_0\}, \\
    n^2 v_n &=& \tilde\bY(\bOne)^{\T} \Lambda_1 \tilde\bY(\bOne) +
    \tilde\bY(\bZero)^{\T} \Lambda_0 \tilde\bY(\bZero) + 
    2 \tilde\bY(\bOne)^{\T} \Lambda_{\tau} \tilde\bY(\bZero).
\end{eqnarray*}
Therefore,
\begin{eqnarray*}
    && n^{2} \left\{v_n - v_n(\beta_1,\beta_0)\right\} \\
    &=& n^2 v_n - \tilde\bY(\bOne)^{\T} \Lambda_1 \tilde\bY(\bOne) -
    \tilde\bY(\bZero)^{\T} \Lambda_0 \tilde\bY(\bZero) - 
    2 \tilde\bY(\bOne)^{\T} \Lambda_{\tau} \tilde\bY(\bZero) \\
    &&- 
    \beta_1^{\T} \bX^{\T} \Lambda_1 \bX \beta_1 -
    \beta_0^{\T} \bX^{\T} \Lambda_0 \bX \beta_0 - 
    2 \beta_1^{\T} \bX^{\T} \Lambda_{\tau} 
    \bX \beta_0 \\
    &&+ \beta_1^{\T}\bX^{\T} \{\Lambda_1\tilde\bY(\bOne)+\Lambda_\tau\tilde\bY(\bZero)\}
    +\beta_0^{\T}\bX^{\T} \{\Lambda_0\tilde\bY(\bZero)+\Lambda_\tau\tilde\bY(\bOne)\} \\
    &=& n^2 L(\beta_1,\beta_0).
\end{eqnarray*}
\qed

\subsection{Proof of Theorem~\ref{thm::cov_adj_clt}}

\noindent \textbf{Parts (a) and (b).}
We prove the two parts together using the following two steps. We denote $\kappa(m) = m^{1+\delta}$ for convenience. 

\noindent \textbf{Step I. Convergence of the regression coefficients.} 

Define the population limit counterpart for the closed-form solution $(\beta_1^\star, \beta_0^\star)^\T$ as
\begin{eqnarray}
\begin{pmatrix}
    \beta_1^\star\\
    \beta_0^\star
\end{pmatrix}
    = \Omega_{xx}^{\dagger}\Omega_{yx}, 
\end{eqnarray}
where 
\begin{align}
    \Omega_{xx} \ =\  
    \begin{pmatrix}
    \Omega_{xx,11} & \Omega_{xx,10} \\
    \Omega_{xx,01} & \Omega_{xx,00}
    \end{pmatrix}, \quad 
    \Omega_{yx} \ =\ 
    \begin{pmatrix}
    \Omega_{yx,11} + \Omega_{yx,01} \\
    \Omega_{yx,00} + \Omega_{yx,10}
    \end{pmatrix}.
\end{align}
By Assumption \ref{assump::limit_values}, we have 
\begin{eqnarray*}
    \frac{1}{\kappa(m)}
    \begin{pmatrix}
    \bX^{\T} \Lambda_1 \bX & \bX^{\T} \Lambda_\tau \bX \\
    \bX^{\T} \Lambda_\tau \bX & \bX^{\T} \Lambda_0 \bX
    \end{pmatrix} 
    \rightarrow 
    \Omega_{xx}. 
\end{eqnarray*}
By similar arguments as in Theorem~\ref{thm::var_est}, under Assumption \ref{assump::limit_values}, the following holds asymptotically in probability:
\begin{eqnarray*}
    \frac{1}{\kappa(m)}
    \begin{pmatrix}
    \sumij \frac{T_iT_j X_i (Y_j - \hat\mu_1) (\Lambda_1)_{i,j}}{p^{|\cGijcup|}} + \sumij \frac{C_iC_j X_i (Y_j - \hat\mu_0) (\Lambda_\tau)_{i,j}}{(1-p)^{|\cGijcup|}} \\
    \sumij \frac{T_iT_j X_i (Y_j - \hat\mu_1) (\Lambda_\tau)_{i,j}}{p^{|\cGijcup|}} + \sumij \frac{C_iC_j X_i (Y_j - \hat\mu_0) (\Lambda_0)_{i,j}}{(1-p)^{|\cGijcup|}}
    \end{pmatrix}^{\T}
    \rightarrow
    \begin{pmatrix}
    \Omega_{yx,11} + \Omega_{yx,01} \\
    \Omega_{yx,00} + \Omega_{yx,10}
    \end{pmatrix}. 
\end{eqnarray*}
Therefore, we conclude that 
\begin{eqnarray}
    (\hat\beta_1, \hat\beta_0)
    -
    (\beta_1^\star, \beta_0^\star) = o_{P}(1). 
\end{eqnarray} 

\noindent \textbf{Step II.
Consistency and asymptotic distribution of $\hat\tau(\hat \beta_1, \hat \beta_0)$. }

For the variance $v_n(\beta_1^\star,\beta_0^\star)$, we have
\begin{eqnarray*}
    \frac{n^2}{\kappa(m)}\cdot v_n(\beta_1^\star,\beta_0^\star) 
    &=&
    \frac{1}{\kappa(m)}(\{\tilde\bY(\bOne) - \bX \beta_1^\star\}^{\T} \Lambda_1\{\tilde\bY(\bOne) - \bX \beta_1^\star \} + \{\tilde\bY(\bZero) - \bX \beta_0^\star\}^{\T} \Lambda_0\{\tilde\bY(\bZero) - \bX \beta_0^\star \} \\
    &&+ 2 \{\tilde\bY(\bOne) - \bX \beta_1^\star\}^{\T} \Lambda_{\tau} 
    \{\tilde\bY(\bZero) - \bX \beta_0^\star\})\\
    &\rightarrow& \Omega_{yy,11} + \Omega_{yy,00} + 2\Omega_{yy,10} + \beta_1^{\star\T}\Omega_{xx,11}\beta_1^\star + \beta_0^{\star\T}\Omega_{xx,00}\beta_0^\star +
    2\beta_1^{\star\T}\Omega_{xx,10}\beta_0^\star \\
    &&-\beta_1^{^\star\T}(\Omega_{xy,11}+\Omega_{xy,10})
    -
    \beta_0^{\star\T}(\Omega_{xy,00}+\Omega_{xy,01}). 
\end{eqnarray*}
This suggests that $v_n(\beta^\star_1, \beta_0^\star) \asymp {\kappa(m)}/{n^2}$.

The difference between $\hat\tau(\hat\beta_1, \hat\beta_0)$ and $\hat{\tau}(\beta_1^\star,\beta_0^\star)$ is
\begin{eqnarray*}
&& \hat\tau(\hat\beta_1, \hat\beta_0) - \hat{\tau}(\beta_1^\star,\beta_0^\star) \\
&=& n^{-1} \sumi \frac{T_i (\hat\beta_1 - \beta_1^\star)^\T X_i}{p^{\G_{+i}}} \Big/ n^{-1}\sumi 
\frac{T_i}{p^{\G_{+i}}}
- 
n^{-1}\sumi \frac{C_i (\hat\beta_0 - \beta_0^\star)^\T X_i}{(1-p)^{\G_{+i}}} \Big/ n^{-1}\sumi \frac{C_i}{(1-p)^{\G_{+i}}}.
\end{eqnarray*}

By the consistency of the optimization solutions
\begin{eqnarray*}
\hat\beta_1 - \beta_1^\star = o_{P}(1), \quad  
\hat\beta_0 - \beta_0^\star = o_{P}(1),
\end{eqnarray*}
and the facts that  
\begin{eqnarray*}
n^{-1} \sumi \frac{T_i X_i}{p^{\G_{+i}}} = O_{P}\left(\frac{\kappa(m)}{n^2}\right), 
\quad 
n^{-1}\sumi \frac{C_i X_i}{(1-p)^{\G_{+i}}}
=
O_{P}\left(\frac{\kappa(m)}{n^2}\right), \\
n^{-1} \sumi \frac{T_i}{p^{\G_{+i}}} = 1 + O_{P}\left(\frac{\kappa(m)}{n^2}\right), 
\quad 
n^{-1}\sumi \frac{C_i }{(1-p)^{\G_{+i}}}
=
1 + O_{P}\left(\frac{\kappa(m)}{n^2}\right),
\end{eqnarray*}
following similar arguments as in proof of Theorem~\ref{thm::asym_dist}, we can conclude that
\begin{eqnarray}
|\hat\tau(\hat\beta_1, \hat\beta_0) - \hat{\tau}(\beta_1^\star,\beta_0^\star)|
&=&
o_{P}\left(\kappa(m)^{1/2}/{n}\right).\label{eqn:tau-diff} \\
\ \ \notag
\end{eqnarray}

Now combining the variance rate $v_n(\beta^\star_1, \beta_0^\star) \asymp {\kappa(m)}/{n^2}$ and \eqref{eqn:tau-diff}, we conclude that
\begin{eqnarray*}
\frac{|\hat\tau(\hat\beta_1, \hat\beta_0) - \hat{\tau}(\beta_1^\star,\beta_0^\star)|}{v_n(\beta_1^\star, \beta_0^\star)^{1/2}}
=
o_{P}\left(1\right).
\end{eqnarray*}
We also have the asymptotic equivalence between $\hat\tau(\hat\beta_1, \hat\beta_0)$ and $\hat\tau(\tilde\beta_1, \tilde\beta_0)$ because $\beta_1^\star$ and $\beta_0^\star$ are the limit values of $\tilde\beta_1$ and $\tilde\beta_0$, respectively, by Assumption~\ref{assump::limit_values}. 

By Proposition~\ref{prop::asym_dist_cov_adj}, $\hat{\tau}(\beta_1^\star, \beta_0^\star)$ converges in probability to $\tau$. Hence $\hat\tau(\hat\beta_1, \hat\beta_0)$ is also consistent to $\tau$. Besides, under the regularity condition $\Dmaxsupp/n = o(m^{-5/6})$, we have that
\begin{align}
    \frac{v_n(\beta_1^\star, \beta_0^\star)}{\sqrt{m} (\Dmaxsupp/n)^2}
    \asymp 
    \frac{{\kappa(m)}/{n^2}}{\sqrt{m} (\Dmaxsupp/n)^2}
    =
    \frac{\kappa(m)}{\sqrt{m}\Dmaxsupp^2},
\end{align}
which diverges to infinity based on the imposed regularity condition. Therefore, the asymptotic distribution of $\hat\tau(\hat\beta_1, \hat\beta_0)$ follows from Slutsky's Theorem and the fact that 
\begin{eqnarray*}
    \left\{v_n(\beta_1^\star, \beta_0^\star)\right\}^{-1/2}\left\{\hat\tau(\hat\beta_1, \hat\beta_0) - \tau\right\} &\to& \cN(0,1) 
\end{eqnarray*}
in distribution.     

\noindent \textbf{Part (c).}
We have
\begin{eqnarray*}
     &&n^{-2}\sumij \frac{T_iT_j(Y_i - \hat\mu_1 - \hat\beta_1^\T X_i)(Y_j - \hat\mu_1 - \hat\beta_1^\T X_j)(\Lambda_1)_{i,j}}{p^{|\cGijcup|}} \\ 
     &=&
     n^{-2}\sumij \frac{T_iT_j\{\tilde Y_i - \beta_1^{\star\T}X_i - (\hat\mu_1 - \mu_1) - (\hat \beta_1 - \beta^\star)^\T X_i\}\{\tilde Y_j - \beta_1^{\star\T}X_j - (\hat\mu_1 - \mu_1) - (\hat \beta_1 - \beta^\star)^\T X_j\}(\Lambda_1)_{i,j}}{p^{|\cGijcup|}} \\ 
     &=& 
     n^{-2}\sumij \frac{T_iT_j(\tilde Y_i - \beta_1^{\star\T}X_i)(\tilde Y_j - \beta_1^{\star\T}X_j)(\Lambda_1)_{i,j}}{p^{|\cGijcup|}} \\ 
     &&+ 2n^{-2}\sumij \frac{T_iT_j(\tilde Y_i - \beta_1^{\star\T}X_i)\{ (\hat\mu_1 - \mu_1) + (\hat \beta_1 - \beta^\star)^\T X_j\}(\Lambda_1)_{i,j}}{p^{|\cGijcup|}}\\
     &&+ n^{-2}\sumij \frac{T_iT_j\{ (\hat\mu_1 - \mu_1) + (\hat \beta_1 - \beta^\star)^\T X_i\}\{(\hat\mu_1 - \mu_1) + (\hat \beta_1 - \beta^\star)^\T X_j\}(\Lambda_1)_{i,j}}{p^{|\cGijcup|}}\\
     &=& \textup{I} + \textup{II} + \textup{III}.
\end{eqnarray*}
Following a similar proof as that of Theorem~\ref{thm::var_est}, we have 
\begin{eqnarray*}
    \textup{I} = v_1^\star(\beta_1^\star)
    +
    O_{P}(n^{-3/2}\Dmaxsupp^{-3/2}), \quad
    \textup{II} = o_{P}(n^{-3/2}\Dmaxsupp^{-3/2}), \quad 
    \textup{III} = o_{P}(n^{-3/2}\Dmaxsupp^{-3/2}). 
\end{eqnarray*}
Meanwhile, due to the fact that $v_1^\star(\beta_1) \asymp n^{-1}\Dmaxsupp$, we have
\begin{eqnarray*}
    \hat v_{1,n}(\hat\beta_1) = {v_1^\star(\beta_1^\star)} + O_{P}(n^{-3/2}\Dmaxsupp^{-3/2})
\end{eqnarray*}
and similarly
\begin{eqnarray*}
    \hat v_{0,n}(\hat\beta_0) = v_0^\star(\beta_0^\star) + O_{P}(n^{-3/2}\Dmaxsupp^{-3/2}). 
\end{eqnarray*}
Therefore, 
\begin{eqnarray*}
    \left\{\hat v_{1,n}(\hat\beta_1)\right\}^{1/2} 
    + 
    \left\{\hat v_{0,n}(\hat\beta_0)\right\}^{1/2} 
    =
    \left\{v_1^\star(\beta_1^\star)\right\}^{1/2}
    +
    \left\{v_0^\star(\beta_0^\star)\right\}^{1/2}
    +
    O_{P}(n^{-3/4}\Dmaxsupp^{-3/4}),
\end{eqnarray*}
and thus 
\begin{eqnarray*}
    \frac{\{\hat v_{1,n}(\hat\beta_1)\}^{1/2} 
    + 
    \{\hat v_{0,n}(\hat\beta_0)\}^{1/2}}
    {\{v_1^\star(\beta_1^\star)\}^{1/2}
    +
    \{v_0^\star(\beta_0^\star)\}^{1/2}} 
    =
    1
    +
    O_{P}(n^{-1/4}\Dmaxsupp^{-1/4})
\end{eqnarray*}
by the fact that $v_1^\star(\beta_1^\star)\asymp n^{-1}\Dmaxsupp$ and $v_0^\star(\beta_0^\star)\asymp n^{-1}\Dmaxsupp$. Therefore, $\hat v_{n,\textsc{ub}}(\hat\beta_1, \hat\beta_0) / v_{n,\textsc{ub}}(\beta_1^\star, \beta_0^\star)$ converges in probability to 1.

The conservativeness of $v^\star_{\textsc{ub}}(\beta_1^\star, \beta_0^\star)$ for the true variance $v^\star(\beta_1^\star, \beta_0^\star)$ can be established similarly to Theorem~\ref{thm::var_est} when no covariates are adjusted. The trick is to take $\bY(\bOne) - \bX\beta_1^\star$ and $\bY(\bZero) - \bX\beta_0^\star$ as pseudo potential outcomes and apply the Cauchy--Schwarz inequality for the covariance. Details are omitted. 
\qed

\section{Positive semi-definiteness of $\Lambda$ and the proof}
\label{sec::psd_lambda}

The variance estimation for the point estimator and the covariate adjustment strategy we proposed in the main paper involves a crucial matrix:
\begin{eqnarray}\label{eqn::Lambda}
\Lambda = 
\begin{pmatrix}
    \Lambda_1 & \Lambda_\tau \\
    \Lambda_\tau & \Lambda_0
\end{pmatrix}
\end{eqnarray}
For example, to justify the existence of a feasible solution to \eqref{eqn::oracle-opt}, we need the matrix $\Lambda$ to be positive-semidefinite. In this section, we prove several results regarding the properties of \eqref{eqn::Lambda}. Define $\Theta = (\theta_{ij})_{i,j\in[n]}$ with $\theta_{ij} = |\cGijcap|$. For each pair $i,j\in[n]$, $\theta_{ij}$ is the number of intervention units connected to both outcome units $i$ and $j$. Theorem~\ref{thm::asym_dist} involves the matrices $\Lambda_1$ and $\Lambda_0$. Moreover, for a function $f(\cdot):\mathbb R \to \mathbb R$, we denote $f([\Theta]) = (f(\theta_{ij}))_{i,j\in[n]}$ as the element-wise mapping of the matrix $\Theta$ with $f$. Therefore, we can write
\begin{eqnarray*}
    \Lambda_1 = p^{-[\Theta]} - 1, \quad \Lambda_0 = (1-p)^{-[\Theta]} - 1.
\end{eqnarray*}
The following proposition summarizes some properties of several important matrices.
\begin{proposition}[Positive semi-definiteness of $p^{-[\Theta]}$ and $(1-p)^{-[\Theta]}$] \label{prop::psd}
We have the following results: 

(a) the matrices $p^{-[\Theta]}$ and $(1-p)^{-[\Theta]}$ are both positive semi-definite matrices.

(b) the following matrix, denoted by $\Lambda$, is positive semi-definite:
\begin{align}
    \begin{pmatrix}
        \Lambda_1 & \Lambda_\tau \\
        \Lambda_\tau & \Lambda_0 
    \end{pmatrix}. 
\end{align}
\end{proposition}

\begin{proof}[Proof of Proposition~\ref{prop::psd} Part (a)]
Note the following binomial decomposition of the elements:
    \begin{align*}
        p^{-\theta_{ij}} 
        = 
        \Big(1 + \frac{1-p}{p}\Big)^{\theta_{ij}}
        =
        1 
        + 
        \frac{1-p}{p} {\theta_{ij} \choose 1}
        +
        \Big(\frac{1-p}{p}\Big)^2 {\theta_{ij} \choose 2} 
        + \dots 
        + \Big(\frac{1-p}{p}\Big)^{\Smaxsupp} {\theta_{ij} \choose \Smaxsupp}.
    \end{align*}
    Hence, we can express $p^{-[\Theta]}$ as
    \begin{align*}
        p^{-[\Theta]} 
        = 
        \Big(1 + \frac{1-p}{p}\Big)^{[\Theta]}
        =
        1 
        + 
        \frac{1-p}{p} {[\Theta] \choose 1}
        +
        \Big(\frac{1-p}{p}\Big)^2 {[\Theta] \choose 2} 
        + \dots 
        + \Big(\frac{1-p}{p}\Big)^{\Smaxsupp} {[\Theta] \choose \Smaxsupp}.
    \end{align*}
    We next prove that 
    \begin{eqnarray*}
        \Theta_k = {[\Theta] \choose k} \text{ is positive semidefinite for $k\in[\Smaxsupp]$.}
    \end{eqnarray*}
        
    (1) For $k = 1$, $\Theta_1 = \Theta$, and $|\cGijcap|$ represents the number of intervention units to which both outcome units $i$ and $j$ are connected. Note that the $i$-th column of $G$, denoted by $g_i\in\mathbb R^m$, indicates the intervention units that $i$ connects to. This suggests $\theta_{ij}$ is also given by $g_i^\T g_j$, and thus 
    \begin{align*}
        \Theta_1 = G^\T G \succeq 0. 
    \end{align*}

    (2) For $k = 2$, we notice the following fact:
    \begin{eqnarray*}
        {\theta_{ij} \choose 2} \text{ indicates the number of pairs of intervention units that outcome units $i$ and $j$ belong to.}
    \end{eqnarray*}
     If we further define a new membership matrix $G^{(2)} \in \mathbb{R}^{n\times m(m-1)/2}$, whose rows are indexed by outcome units and columns by pairs of $m$ intervention units, with entry $G^{(2)}_{li}$ representing whether outcome unit $i$ connects to the $l$-th pair. Then we can show that 
    \begin{align*}
        \Theta_2 = G^{(2)\T} G^{(2)} \succeq 0. 
    \end{align*}

    Similarly, we can extend the above trick to $\Theta_k$. We define $G^{(k)}$ as the membership matrix indicating whether outcome unit $i$ connects to a $k$-tuple of the intervention units. Then we can show
    \begin{align*}
        \Theta_k = G^{(k)\T} G^{(k)} \succeq 0. 
    \end{align*}
    This concludes the claim. 
\end{proof}

\begin{proof}[Proof of Proposition~\ref{prop::psd} Part (b)]
For symmetry of presentation, denote $G^{(1)} = G$. Consider any vectors $\alpha \in \mathbb R^n$ and $\beta \in \mathbb R^n$, we have
\begin{align*}
\alpha^\T \left( p^{[-\Theta_1]}-1 \right) \alpha = \left(\frac{1-p}{p}\right) \alpha^\T  G^{(1)\T} G^{(1)}  \alpha + \left(\frac{1-p}{p}\right)^2 \alpha^\T  G^{(2)\T} G^{(2)}  \alpha + \cdots, 
\end{align*}
where 
\begin{align*}
\alpha^\T G^{(s)\T} = \left( \sum_{i \in \cG_{k_1+}\cap \cdots \cap \cG_{k_s+}} \alpha_i \right)_{1\le k_1<\dots<k_s\le m}. 
\end{align*}
Same for the $\beta^\T \left( (1-p)^{-[\Theta_1]}-1 \right) \beta$ part. 
Now 
\begin{align*}
\alpha^\T \Theta_1 \beta = \sumi \sumj \alpha_i \beta_j \indicator\{(\Theta_1)_{ij} \neq 0\}. 
\end{align*}

Note that for any index set $\mathcal{J}$, 
\begin{align*}
2 \left| \left( \sum_{i \in \mathcal{J}} \alpha_i \right) \left( \sum_{j \in \mathcal{J}} \beta_j \right) \right| 
&= 2 \left| \sqrt{\left(\frac{1-p}{p}\right)^k} \left( \sum_{i \in \mathcal{J}} \alpha_i \right) \cdot \sqrt{\left(\frac{p}{1-p}\right)^k} \left( \sum_{j \in \mathcal{J}} \beta_j \right) \right| \\
&\leq \left(\frac{1-p}{p}\right)^k \left( \sum_{i \in \mathcal{J}} \alpha_i \right)^2 + \left(\frac{p}{1-p}\right)^k \left( \sum_{j \in \mathcal{J}} \beta_j \right)^2
\end{align*}

Therefore, for $i, j$, $(\Theta_1)_{ij} \neq 0$, we do the following:

\noindent(1) First, we have
\begin{eqnarray*}
    &&\left(\frac{1-p}{p}\right) \alpha^\T G^{(1)\T} G^{(1)} \alpha 
    + 
    \left(\frac{p}{1-p}\right) \beta^\T G^{(1)\T} G^{(1)} \beta \\
    & = &
    \left(\frac{1-p}{p}\right)
    \sumk \left(\sum_{i\in\cG_{k+}}\alpha_i\right)^2
    +
    \left(\frac{p}{1-p}\right)\sum_{k=1}^m
    \left(\sum_{j\in\cG_{k+}}\beta_j\right)^2\\
    & \ge &
    -\sumk \left( \sum_{i,j \in \cG_{k+}} \alpha_i\beta_j \right) 
    =  
    -\sumij {\theta_{ij} \choose 1}\alpha_i\beta_j \indicator\{\cGijcap \neq \varnothing\}. 
\end{eqnarray*}

% we get the lower bound 
% \begin{align*}
% - {\theta_{ij}\choose 1} \alpha_i \beta_j;
% \end{align*}

\noindent(2) Second, we have
\begin{eqnarray*}
    && \left(\frac{1-p}{p}\right)^2 \alpha^\T G^{(2)\T} G^{(2)} \alpha 
    + 
    \left(\frac{p}{1-p}\right)^2 \beta^\T G^{(2)\T} G^{(2)} \beta\\
    & = & 
    \left(\frac{1-p}{p}\right)^2
    \sum_{(kl)\subset[m]} \left(\sum_{i\in\cG_{k+}\cap\cG_{l+}}\alpha_i\right)^2
    +
    \left(\frac{p}{1-p}\right)^2
    \sum_{(kl)\subset[m]}
    \left(\sum_{j\in\cG_{k+}\cap\cG_{l+}}\beta_j\right)^2\\
    & \ge &
    \sum_{(kl)\subset[m]} \left( \sum_{i,j \in \cG_{k+}\cap\cG_{l+}} \alpha_i\beta_j \right)
    = \sumij {\theta_{ij} \choose 2}\alpha_i\beta_j \indicator\{\cGijcap \neq \varnothing\}.
\end{eqnarray*}

\noindent(3) Similarly, we have 
\begin{align*}
    \left(\frac{1-p}{p}\right)^3 \alpha^\T G^{(3)\T} G^{(3)} \alpha 
    + 
    \left(\frac{p}{1-p}\right)^3 \beta^\T G^{(3)\T} G^{(3)} \beta 
    \ge -\sumij {\theta_{ij} \choose 3}\alpha_i\beta_j \indicator\{\cGijcap \neq \varnothing\}.
\end{align*}

This process can be conducted until $k = \Smaxsupp$. Hence, taking an addition, we can obtain
\begin{align*}
    \left\{-{\theta_{ij}\choose 1} + {\theta_{ij}\choose 2} -{\theta_{ij}\choose 3} + \dots\right\} \alpha_i \beta_j 
    = 
    -\alpha_i \beta_j + (1-1)^{\theta_{ij}}\alpha_i \beta_j
    =
    -\alpha_i \beta_j. 
\end{align*}
This concludes the proof. 
\end{proof}

\end{document}